\documentclass[a4paper]{article}
\usepackage[]{amsmath}
\usepackage{authblk}
\usepackage{bm}
\usepackage{physics}
\usepackage{multirow}
\usepackage{glossaries}
\usepackage[]{hyperref}
\usepackage[]{graphicx}
\usepackage{array}
\usepackage[]{hyperref}
\usepackage{caption}
\usepackage{subcaption}
\usepackage[]{color}
\usepackage{soul}
\usepackage[margin=1in]{geometry}
\usepackage{amssymb}

\title{Inferring proximity from \glsdesc{ble} \acrshort{rssi} with \glsdesc{uks}s}
\author[1]{Tom Lovett}
\author[1, *]{Mark Briers}
\author[1]{Marcos Charalambides}
\author[1]{Radka Jersakova}
\author[1]{James Lomax}
\author[1, 2]{Chris Holmes}

\affil[1]{\small The Alan Turing Institute, London, U.K.}
\affil[2]{\small University of Oxford, Oxford, U.K.}
\affil[*]{\small Corresponding author: Mark Briers, mbriers@turing.ac.uk}

\newacronym{rssi}{RSSI}{Received Signal Strength Indicator}
\newacronym{ble}{BLE}{Bluetooth Low Energy}
\newacronym{hmc}{HMC}{Hamiltonian Monte Carlo}
\newacronym{nuts}{NUTS}{No U-Turn Sampler}
\newacronym{rmse}{RMSE}{Root Mean-Squared Error}
\newacronym[longplural={Unscented Kalman Smoothers}]{uks}{UKS}{Unscented Kalman Smoother}
\newacronym{roc}{ROC}{Receiver Operating Characteristic}
\newacronym{auc}{AUC}{Area Under Curve}

\DeclareMathOperator*{\argmin}{argmin}

\begin{document}
\maketitle

\begin{abstract}
  The Covid-19 pandemic has resulted in a variety of approaches for managing infection outbreaks in international populations. One example is mobile phone applications, which attempt to alert infected individuals and their contacts by automatically inferring two key components of infection risk: the proximity to an individual who may be infected, and the duration of proximity. The former component, proximity, relies on \acrfull{ble} \acrfull{rssi} as a distance sensor, and this has been shown to be problematic; not least because of unpredictable variations caused by different device types, device location on-body, device orientation, the local environment and the general noise associated with radio frequency propagation. In this paper, we present an approach that infers posterior probabilities over distance given \emph{sequences} of \gls{rssi} values. Using a single-dimensional \acrfull{uks} for non-linear state space modelling, we outline several Gaussian process observation transforms, including: a generative model that directly captures sources of variation; and a discriminative model that learns a suitable observation function from training data using both distance and infection risk as optimisation objective functions. Our results show that good risk prediction can be achieved in $\mathcal{O}(n)$ time on real-world data sets, with the \gls{uks} outperforming more traditional classification methods learned from the same training data.
\end{abstract}

\section{Introduction}
There has been recent global interest in the use of \acrfull{ble} \acrfull{rssi} as a proximity sensor. This is motivated by the international Covid-19 pandemic and the use of mobile phone applications to help control infection propagation through the population. At the time of writing, many of these applications are using \gls{ble} to infer whether people are close together for prolonged periods of time, since proximate, prolonged exposure to an infected person correlates with the probability of infection \cite{ferretti2020, sohrabi2020world}.

Unfortunately, \gls{ble} \gls{rssi} is a very noisy sensor of proximity. Due to the usual vagaries of radio frequency propagation in general environments, e.g. multipath, reflection, shadowing and fading, it becomes challenging to infer proximity from observed values without taking into account uncertainties in the data generating process.

\begin{figure}[tb]
  \centering
  \begin{subfigure}{.49\textwidth}
    \includegraphics[width=\textwidth, trim={1.5cm 0 1cm 0}, clip]{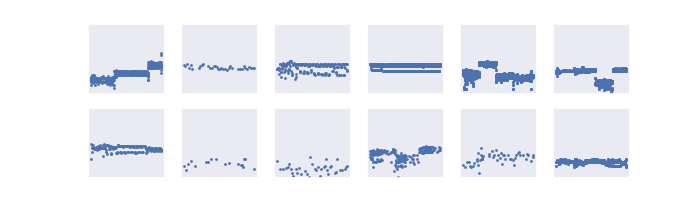}
    \caption{Trinity College Dublin data, from \cite{leith2020coronavirus}.}
  \end{subfigure}
  \begin{subfigure}{.49\textwidth}
    \includegraphics[width=\textwidth, trim={1.5cm 0 1cm 0}, clip]{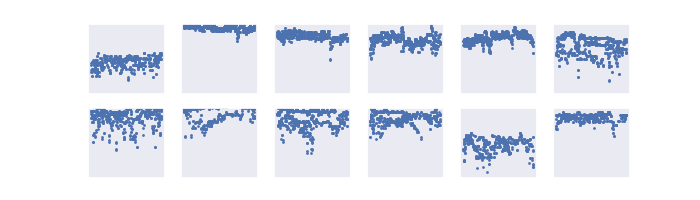}
    \caption{MIT PACT data \cite{pact}.}
  \end{subfigure}
  \caption{Samples of \gls{rssi} data over time from two sources. The Trinity College data here are all recorded at a proximity of $1$m, and the MIT data at $6$ft. All plots across both sources share the same $y$-axes $\left[ -100, -50 \right]$.}
  \label{fig:rssi}
\end{figure}

To illustrate the highly variable behaviour of \gls{rssi} values, consider the plots in Figure \ref{fig:rssi}. These all show \gls{rssi} values recorded at a fixed distance ($1$m for the Trinity College data sets, and $6$ft for the MIT data sets) over time. Notice the extreme shifts and large variances; these are due to a multitude of sources, e.g. device type, orientation, position on body and the person's local environment.

There have been various attempts to work with \gls{rssi} in a principled manner in Covid-19 mobile applications. In the popular Exposure Notification system \cite{appleapi, googleapi} used by Google and Apple devices, \gls{rssi} is ``discretised'' into buckets, and bucket thresholds are set by application developers. The documentation \cite{googleapible} reasons that \gls{rssi} is useful for inferring close proximity, but not at all effective for inferring larger distances.

In this paper, we present a probabilistic model for inferring proximity using \gls{ble} \gls{rssi} observations. This model uses a \acrfull{uks} \cite{julier1997new, julier1995new}, which takes advantage of sequential \gls{rssi} data, and models multiple sources of uncertainty in the data distribution variance. For the data distribution, we use a Gaussian process that maps distance to observations, and present two general forms: a generative model, which directly models the sources of shifts and variance in the observations; and a discriminative model, which learns a suitable observation model from training data. Setting this paper's contribution in context, the estimation of a distribution over proximity is just one technical problem that needs to be addressed; the reader is also referred to \cite{morley2020ethical} for a broader discussion around the ethical considerations around contact tracing app development.

\section{Related work}
Since the Covid-19 outbreak, there has been a surge of interest in proximity inference using \gls{ble} \gls{rssi} data. Since risk of infection is a function of, \emph{iter alia}, proximity between individuals and the duration of contact \cite{ferretti2020, briers2020risk}, automatically inferring these two properties for any contact ``event'' is crucial.

Perhaps the most notable work exploring the effects of \gls{ble} \gls{rssi} in real-world environments, and its potential as a sensor of proximity, are the series of papers by Trinity College Dublin. For example, in \cite{leith2020coronavirus}, the authors demonstrate the extreme shifts and variation in \gls{rssi} over a variety of contexts, including on public transport, in supermarkets, walking in public streets and sitting at desks. There are also scenarios demonstrating the shifts in \gls{rssi} from simply putting a mobile device in a pocket or a bag. The data in Figure \ref{fig:rssi} are taken from this paper.

In \cite{leith2020measurement}, the authors show that in this public transport environment with 60 device pairs, the Exposure Notification system with particular parameter settings did not detect genuine contact events, though minor improvements were made through parameter variation. Further evidence of the difficulties of using \gls{rssi} is illustrated in \cite{leith2020tram}, with the best performing parameter settings achieving the equivalent of random selection.

Other recent studies into Bluetooth for proximity detection include: \cite{liu2013face}, where further tests of Bluetooth in indoor and outdoor environments, as well as device concealment variations, show the volatility of \gls{rssi} and the ambiguity at larger distances; and \cite{montanari2017study}, where \gls{ble} is used for proximity detection in the workplace, albeit with exposed bracelets and additional environment sensors, and good binary classification of contact events is achieved through varying scan windows at a cost of power consumption (the drop in performance for low-power, lower frequency methods even in a ``good coverage'' environment is noted).

Methods that make use of \gls{rssi} sequences include \cite{rodas2008bayesian}, which uses particle filtering on Bluetooth \gls{rssi} to infer proximity in idealised environments. The findings are unlikely to translate to everyday mobile phone use however, since they were obtained from sensor networks with different Bluetooth hardware, where the setup was designed specifically for object tracking. Other methods have used Kalman filters, which assume linear transforms for state transition and observation. In \cite{subhan2013kalman}, a Kalman filter was applied to the indoor location estimation problem but, again, the hardware and experiment setup used (high-power class 1 Bluetooth sensor networks set up for tracking) do not translate to real-world mobile device use. Similar studies used augmented sensors, e.g. an inertial sensor \cite{yoon2015}, in environments designed for tracking, e.g. \cite{zhou2017bluetooth} and some have considered particle filters with Gaussian processes for sequential modelling \cite{jadidi2017gaussian}.

In direct response to the Covid-19 crisis, other approaches to proximity detection with realistic context and \gls{ble} hardware have been studied. These include Gorce, Egan and Gribonval \cite{gorce2020}, who use calibrated \gls{ble} \gls{rssi}, where shifts due to device type, position and environment are considered, as well as probabilistic modelling of fading and shadowing. The authors then use Bayesian inference to compute a posterior distribution over distance given (averaged, calibrated) \gls{rssi} observations. This posterior uses a constrained uniform prior with Gaussian data distribution over the \gls{rssi} observations. Different estimators are derived from this posterior, such as the maximum \emph{a posteriori} and maximum likelihood distance. These are then used in various risk scoring approaches, and experiments show reasonable risk inference using this method. This work is similar to ours in its Bayesian approach, though it does not take sequences into account. Moreover, we model shifts as random variables in the data distribution rather than through correction and averaging.

\section{Posterior proximity inference}
Given a sequence of observed \gls{rssi} random variables $R_1,\dots,R_{T'}$, we wish to compute the probability distribution over device proximity, or distance, at each observation, i.e. $D_1,\dots,D_{T'}$. Since \gls{rssi} data are likely to be aperiodic, bursty and unreliable, we might wish to treat $D_1,\dots,D_{T'}$ as a subsequence of a larger, periodic sequence $D_1,\dots,D_T$, where $T' \le T$ and also infer the proximity at points where observations are not present (see Section \ref{sec:imputation}).

Since (negative) \gls{rssi} appears to have a longer tail to $\infty$ in empirical data, e.g. Figure \ref{fig:rssi} and the plots in \cite{gorce2020}, and that \gls{ble} typically has transmission power $\le 0$dBm, we model $-R$ as a log-normal random variable. This is in contrast to popular radio propagation models such as the log-distance path model, which assumes the variability in $R$ follows a Gaussian distribution. For completeness, we have replicated all results in this paper under the assumption of $R$ being a Gaussian random variable (see Appendix \ref{sec:appendix}), and the log-normal model appears to be more resilient to fluctuations in \gls{rssi}, as discussed in Section \ref{sec:discussion}.

Under this log-normal model, we are interested in modelling the data distribution of the Gaussian random variable $X = \log(-R) \in \mathbb{R}$ conditioned on a distance (or proximity) variable $D \in (0, \infty)$, where $R \in (-\infty, 0)$ is \gls{rssi} in dBm at distance $d$\footnote{We assume \gls{ble} transmission power is always $< 0$dBm}. We assume this conditional distribution is characterised by a collection of $D$-dependent parameters $\Theta(D)$,
\begin{equation*}
  F_{X \mid D}\left( X \mid D, \Theta(D) \right),
\end{equation*}
and, given this data distribution for $X \mid D$ along with the sequence of observations $X_1,\dots,X_{T'}$, our goal is to infer the posterior distribution over $D_t$ at each time index $t \in \{1,\dots,T\}$
\begin{equation*}
  F_{D_t \mid X_1,\dots,X_{T'}} \left( D_t \mid X_1,\dots,X_{T'}, \Theta(D_t) \right).
\end{equation*}

We further assume that this distribution (and the data distribution) admits a density with respect to Lebesgue measure, and we compute
\begin{equation*}
  p_{D_t \mid X_1,\dots,X_{T'}}\left( D_t \mid X_1,\dots,X_{T'}, \Theta(D_t) \right) \propto p_{X_1,\dots,X_{T'} \mid D_t}\left( X_1,\dots,X_{T'} \mid D_t, \Theta(D_t) \right)p_{D_t}\left( D_t \right).
\end{equation*}

This is a classic problem in dynamical systems' theory, and is well suited to methods such as Kalman filtering, smoothing and other derivatives. The choice of method depends on the form of the data distribution, and performance depends heavily on the quality of the data distribution, i.e. how closely it matches nature's ``true'' distribution.

\subsection{Unscented Kalman filtering and smoothing}
The Kalman filter and smoother are classic methods for performing posterior inference over latent variables $\mathbf{x}_1,\dots,\mathbf{x}_T \in \mathbb{R}^m$ given discrete sequences of observed vectors in $\mathbf{z}_1,\dots,\mathbf{z}_T \in \mathbb{R}^n$. The traditional Kalman smoother assumes that the latent sequence has the Markov property, all transforms are linear and that all stochastic sources are Gaussian. The state transition model, assuming no control inputs, is
\begin{equation}
  \mathbf{x}_{t + 1} = \mathbf{A}_{t + 1} \mathbf{x}_t + \mathbf{w}_{t + 1},
  \label{eq:kf}
\end{equation}
where $\mathbf{w}_t \sim \mathcal{N}\left( \mathbf{0}, \mathbf{Q}_t \right)$. The observation model is
\begin{equation}
  \mathbf{z}_t = \mathbf{B}_t \mathbf{x}_t + \mathbf{v}_t,
  \label{eq:kfo}
\end{equation}
where $\mathbf{v}_t \sim \mathcal{N}\left( \mathbf{0}, \mathbf{R}_t \right)$.

Unfortunately, in our application of inferring proximity from \gls{ble} \gls{rssi}, the observation model is non-linear and the latent variables $D$ only have non-negative support. We can work around the latter problem by assuming $D_t \in \mathbb{R}$ and transforming $D_t$ to its absolute value, i.e. $\lvert D_t \rvert$ -- this implicitly assumes that the transition distribution is \emph{folded} normal, rather than normal. The nonlinearity of the observation model leads us to use an extension to the traditional Kalman smoother: the \acrfull{uks} \cite{julier1997new, wan2000unscented,briers2010smoothing}. The \gls{uks} uses deterministic inspired sampling to allow nonlinear transforms in the model. Our transition model for the \gls{uks} is
\begin{equation}
  D_{t + 1} = \lvert D_t + w_{t + 1} \rvert,
  \label{eq:uks}
\end{equation}
where $w_t \sim \mathcal{N}(0, q_t)$. This model is equivalent to assuming that the two devices are each performing an independent Gaussian random walk, and that the relative proximity transition follows a folded normal distribution. The key parameter in this model is $q$, the variance of the change in proximity between time steps. The observation model is
\begin{equation}
  X_t = \mu\left( D_t \right) + v_t,
  \label{eq:ukso}
\end{equation}
where $v_t \sim \mathcal{N}(0, r_t)$. In other words, $X_t \sim \mathcal{N}\left( \mu\left( D_t \right), r_t \right)$. 

\subsubsection{Posterior imputation}
\label{sec:imputation}
One of the key advantages of the \gls{uks}, and dynamical systems in general, is the ability to infer the posterior distribution over $D_t$, even when an observation $X_t$ does not exist. This is well suited to \gls{rssi} data, which are bursty, aperiodic and unreliable. This illustrates another benefit of posterior inference over sequences of observations, rather than single observations independently. Other approaches use averaging to smooth observations and inferred values over time windows, e.g. \cite{gorce2020}, but the sequential nature of the \gls{uks} allows for more principled imputation, where observations either side of a ``gap'' induce a more realistic trend in the inferred values.

\subsection{Choosing a suitable data distribution}
Assuming the data distribution
\begin{equation*}
  F_{X \mid D}\left( X \mid D, \Theta(D) \right)
\end{equation*}
is Gaussian, i.e.
\begin{equation}
  X \mid D \sim \mathcal{N}\left( \mu\left( D; \theta_\mu \right), r\left( D; \theta_r \right) \right),
  \label{eq:dataform}
\end{equation}
with $\{ \theta_\mu, \theta_r \} \in \Theta(D)$ then, for distances $d$,
\begin{equation*}
  \left( X_d \right), d \in (0, \infty),
\end{equation*}
is a Gaussian process and, for any given $d$, the Gaussian for $X_d$ depends entirely on the  hyperparameters $\Theta(D)$. Thus, we can encode knowledge of the distribution of $X$ at certain distances in our choices for the hyperparameters. If we have access to appropriate training data, we can use these data to learn $\Theta(D)$ for \gls{rssi} behaviour in general environments. There are two main approaches for doing this: a discriminative approach, which learns representative parameters from training data; and a generative approach, which models directly the sources of \gls{rssi} variability. We consider both approaches and assess their performance in subsequent sections.

\section{Data distribution form}
In this section, we outline a model for the Gaussian data distribution
\begin{equation*}
  F_{X \mid D}\left( X \mid D, \Theta(D) \right).
\end{equation*}
Unfortunately, the vagaries of radio frequency propagation within different environments make physical modelling of this distribution very difficult. 

Empirical data, e.g. \cite{h0h1, gorce2020, pact, leith2020coronavirus} suggest that the distribution of $R$ in a ``clean'' environment, e.g. an anechoic chamber, has a unimodal, asymmetric form, with a long tail towards $-\infty$. We assume that the distribution of $X$ is unimodal and symmetric about the mode, and a member of the location-scale family of distributions.

For a given distance $D=d$, we assume the existence of a fixed function $f: (0, \infty) \to \mathbb{R}$, which captures the physics of radio propagation in free space as a function of distance. We use a simple \gls{rssi} propagation model, which approximates line-of-sight received power in free space using the Friis transmission equation,
\begin{equation*}
    P_r = P_t G_t G_r \left(\frac{\lambda}{4 \pi d}\right)^2,
\end{equation*}
where $P_r$ is received power (in W); $P_t$ is transmitted power; $G_t$ and $G_r$ are transmitter and receiver gains respectively; $d$ is distance between transmitter and receiver (in m); and $\lambda$ is wavelength (in m). We use the decibel conversion
\begin{equation}
  g(d) = 20 \log_{10}\left( \frac{\lambda}{4\pi d} \right),
  \label{eq:friis}
\end{equation}
where $\lambda=0.125$ is Bluetooth wavelength in metres\footnote{This is for the $2402$MHz advertising channel. Future work may wish to also consider the $2426$MHz and $2480$MHz channels, which equate to $0.123$m and $0.121$m wavelengths respectively.}. We assume the transmitted power to be $0$dBm, and that antenna gains for the transmitter and receiver are captured in the shift variables below. The base function $f(d)$ is then
\begin{equation}
  f(g(d)) = \log(-g).
  \label{eq:base_function}
\end{equation}

In our model, this function can be shifted by a finite number $N_s$ of independent random variables $Y_i \mid d$ -- which may represent, e.g. antenna orientations, device model differences and changes in the physical environment -- plus some zero-mean unattributable, independent, distance-invariant noise $Z$. With these forms and assumptions, we have, given $D=d$
\begin{equation*}
  X \mid d = f(d) + \sum_{i=1}^{N_s} Y_i \mid d + Z,
\end{equation*}
and, 
\begin{align}
  \mathbb{E}\left[ X \mid D = d \right] &= f(d) + \sum_{i=1}^{N_s} \mathbb{E}\left[ Y_i \mid d \right] + \mathbb{E}\left[ Z \right],\nonumber \\
  &= f(d) + \sum_{i=1}^{N_s} \int_\mathbb{R} y_i F_{Y_i \mid d}\left( \dd{y_i} \right), 
  \label{eq:exp}
\end{align}
with
\begin{align}
  \operatorname{Var}\left( X \mid D=d \right) &= \sum_{i=1}^{N_s} \operatorname{Var}\left( Y_i \mid d \right) + \operatorname{Var}\left( Z \right), \nonumber \\
  &= \sum_{i=1}^{N_s} \int_\mathbb{R} \left( y_i - \mathbb{E}\left[ Y_i \mid d \right] \right)^2 F_{Y_i \mid d}\left( \dd{y_i} \right) + \int_\mathbb{R} z^2 F_Z\left( \dd{z} \right).
  \label{eq:var}
\end{align}

\section{Generative model}
In this section, we outline a generative model for certain shift variables $Y_i$ and noise variable $Z$, each of which we assume to have Gaussian mixture form with $K_i$ components at distance $d$,
\begin{equation}
  Y_i \mid d \sim \sum_{k = 1}^{K_i} \pi_k \mathcal{N}\left( \mu_k, \sigma_k^2 \right),
  \label{eq:gmm}
\end{equation}
and  -- in general -- unknown $d$-specific $\pi_k, \mu_k$ and $\sigma_k^2$. For each variable $Y_i$, we assume we have access to some empirical observation data for each component $k$: $\mathcal{D}_k$ (which could be empty, i.e. $\mathcal{D}_k = \varnothing$). We place a conjugate normal-inverse-gamma prior over each Gaussian component's parameters $\mu_k, \sigma_k^2$ to obtain the posterior distribution,
\begin{equation*}
  \mu_k, \sigma_k^2 \mid \mathcal{D}_k, \theta_k \sim \operatorname{NIG}\left(m_k, \lambda_k, \alpha_k, \beta_k  \right),
\end{equation*}
and a conjugate Dirichlet prior over the mixture components, to obtain the posterior
\begin{equation*}
  \pi_k \mid \mathcal{D}_k, \theta_k \sim \operatorname{Dirichlet}(\bm{\alpha}),
\end{equation*}
and marginalise over the unknown parameters in Equation \ref{eq:gmm} to obtain the posterior predictive distribution
\begin{equation}
  p_{Y_i \mid d}\left( Y_i \mid \mathcal{D}_1,\dots,\mathcal{D}_{K_i}, \Theta_i \right) = \int_{\bm{\pi}} \sum_{k=1}^{K_i} \pi_k p_{Y_i \mid d}\left( Y_i \mid \mathcal{D}_k, \theta_k \right) \dd{\bm{\pi}},
  \label{eq:mixt}
\end{equation}
where each component has the form
\begin{align}
  p_{Y_i \mid d}\left( Y_i \mid \mathcal{D}_k, \theta_k \right) &= \int_{-\infty}^\infty \int_0^\infty  p\left( Y_i \mid m, s \right) p\left( m, s \mid \mathcal{D}_k, \theta_k \right) \dd{s} \dd{m}, \nonumber \\
  &= t_{2\alpha_k}\left(Y_i \mid m_k, \left[\frac{(1 + \lambda_k)\beta_k}{\lambda_k \alpha_k}\right]^{\frac{1}{2}} \right),
  \label{eq:dtmarg}
\end{align}
i.e. each component is $t_\nu\left( \cdot \mid \mu, \sigma \right)$: a non-standard Student's $t$-distribution with $\nu$ degrees of freedom. Thus our concrete distribution for $Y_i \mid d$ is an average mixture of Student's $t$-distributions. Using standard results from conjugacy and $\{y_1\dots,y_{N_k}\} \in \mathcal{D}_k$, with $\bar{y} := N_k^{-1} \sum_i y_i$,
\begin{align}
  m_k &= \frac{\lambda_0 \mu_0 + \sum_i y_i}{\lambda_0 + N_k},\nonumber \\
  \lambda_k &= \lambda_0 + N_k,\nonumber \\
  \alpha_k &= \alpha_0 + \frac{N_k}{2},\nonumber \\
  \beta_k &= \beta_0 + \frac{1}{2}\left[ \sum_i \left( y_i - \bar{y} \right)^2 + \frac{N_K \lambda_0}{\lambda_0 + N_k} \left( \bar{y} - \mu_0 \right)^2 \right],
  \label{eq:ppparams}
\end{align}
and, for the mixture components, given any multinomial observations of component frequencies $\mathbf{x}_1,\dots,\mathbf{x}_{N_{\bm{\alpha}}}$ with $x_{i, 1} \in \mathcal{D}_1,\dots,x_{i, K} \in \mathcal{D}_K$
\begin{equation}
  \bm{\alpha} = \bm{\alpha}_0 + \sum_{i=1}^{N_{\bm{\alpha}}} \mathbf{x}_i.
  \label{eq:alpha}
\end{equation}

\subsection{Computing $\operatorname{Var}\left( Z \right)$}
For the unattributable noise $Z$, we assume zero-mean Gaussian with unknown variance $\sigma_Z^2$. With zero-mean and a single Gaussian component, the derivation of the previous section shows that
\begin{equation*}
  p_Z\left( Z \mid \theta_Z\right) = t_{2\alpha}( Z \mid 0, \sqrt{\beta/\alpha}),
\end{equation*}
so that
\begin{equation}
  \operatorname{Var}\left( Z \mid \theta_Z \right) = \frac{\beta}{\alpha - 1},
  \label{eq:varz}
\end{equation}
and we require $\alpha > 1$.

\subsection{$Y$ variables}
We focus on three classes of $Y$: device type shifts caused by differences in mobile device hardware; shifts caused by antenna gain variations; and context shifts caused by device usage, e.g. position, location and environment.

\begin{itemize}
  \item \textbf{Device type}: different device types affect shifts and variance of $X$ \cite{gorce2020}, so this is one example of a shift variable $Y$. It also appears that this shift is different for different sender/receiver pairs. Given $N$ device types, i.e. specific makes and models, we have $K = N^2$ sender-receiver pairs. Following Equation \ref{eq:mixt}, our distribution for $Y_i \mid d$ is
  \begin{equation}
    p_{Y \mid d}\left( Y \mid \mathcal{D}_1,\dots,\mathcal{D}_{K}, \Theta \right) = \int_{\bm{\pi}} \sum_{k=1}^K \pi_k p_{Y \mid d}\left( Y \mid \mathcal{D}_k, \theta_k \right) \dd{\bm{\pi}}.
    \label{eq:dshift}
  \end{equation}
  Given choices for prior hyperparameters $\theta_k$, we can collect data on sender/receiver device shifts by measuring them empirically, e.g. in an anechoic chamber, and updating the posterior hyperparameters. We can also use mobile device market share data or survey data to update the Dirichlet parameter in Equation \ref{eq:alpha}.

\item \textbf{Antenna gain}: the Friis transmission equation in Equation \ref{eq:friis} usually includes terms for losses in received power due to directivity of the transmitter and receiver. We encode these losses in a random shift variable $Y$, with a single component. Thus Equation \ref{eq:mixt} becomes
  \begin{equation*}
    p_{Y \mid d}\left( Y \mid \mathcal{D}, \theta \right) = t_{2\alpha}\left(Y \mid m, \left[\frac{(1 + \lambda)\beta}{\lambda \alpha}\right]^{\frac{1}{2}} \right),
  \end{equation*}
  and empirical data on directivity shifts can be use to update the posterior parameters.

\item \textbf{Device position, location and environment}: other sources of variability for $X$ include device (antenna) position, e.g. orientation; device location, e.g. in pocket; and environment, e.g. indoors. Since these are typically not independent, the shift $Y$ depends on the joint distribution of the variables $P$ (position), $L$ (location) and $E$ (environment). If we assume the position, location and orientation variables take values in finite sets, then our mixture model will have $K = N_P \times N_L \times N_E$ components and we have the form of the posterior predictive mixture of Student's $t$-distributions in Equation \ref{eq:mixt}. Again, we can use empirical data for each component $\mathcal{D}_k$ to update the posterior hyperparameters.
\end{itemize}

\section{Discriminative model}
The purpose of a discriminative model is to learn suitable parameters $\Theta(D)$ from training data. This learning process is an optimisation problem, and we wish to find suitable parameters that maximise a particular objective function.

\subsection{Model form and parameters}
There are a range of parametric forms that we could use for the discriminative model of the data distribution in Equation \ref{eq:dataform}. We consider two here: the first is a scaled and shifted base function $f(d)$ with $d$-invariant scale and shift parameters $\theta_{\mu_1}$ and $\theta_{\mu_2}$ respectively, i.e.
\begin{equation}
  \mu\left( d; \theta_\mu \right) = \theta_{\mu_1}f(d) + \theta_{\mu_2},
  \label{eq:shift_model}
\end{equation}
and the second disregards the base function $f$ and assumes a logarithmic form with $d$-invariant scale and intercept, i.e.
\begin{equation}
  \mu\left( d; \theta_\mu \right) = \theta_{\mu_1}\log(d) + \theta_{\mu_2}.
  \label{eq:log_model}
\end{equation}

For both forms, we also have the observation variance $\theta_r$ and transition variance $q$ for the \gls{uks}. The parameters for optimisation are therefore $\Theta(D) = \left\{ \theta_{\mu_1}, \theta_{\mu_2}, \theta_r, q \right\}$.

\subsection{Proximity optimisation}
The first objective function is to minimise the expected average mean-squared error between true distances $D_1,\dots,D_T$ and the expected value of the posterior distribution returned by the \gls{uks} inference process. That is, given $N$ training data sets $\mathcal{D}_1,\dots,\mathcal{D}_N$, where -- to account for missing observations -- $\mathcal{D}_n$ contains a periodic sequence of true distances $(d_1,\dots,d_{T_n})$ and a generally aperiodic, \emph{subsequence} of observations $(x_1,\dots,x_{T_n'})$ with $T_n' \le T_n$. We wish to find
\begin{equation}
  \hat{\Theta} = \argmin_\Theta \mathbb{E}_{\mathcal{D}} \left[ \frac{1}{T_n} \sum_{t = 1}^{T_n} \left( d_t - \mathbb{E} \left[ D_t \mid x_1,\dots,x_{T_n'}, \Theta \right] \right)^2 \right].
  \label{eq:proxopt}
\end{equation}

\subsection{Risk error optimisation}
The second objective function is to minimise the risk error. We use the risk score from \cite{briers2020risk} for one time step under the assumption of maximum infectiousness and minimum time decay as
\begin{equation}
  \gamma(d_t) = \frac{\Delta t}{60} \min\left( 1, \frac{1}{d_t^2} \right),
  \label{eq:risk}
\end{equation}
where $\Delta t$ is the time step (in seconds) between periodic true distances, and search for parameters that minimise the expected average mean-squared risk error,
\begin{equation}
  \hat{\Theta} = \argmin_\Theta \mathbb{E}_{\mathcal{D}} \left[ \frac{1}{T_n} \sum_{t = 1}^{T_n} \left( \gamma\left(d_t\right) - \gamma\left(\mathbb{E}\left[ D_t \mid x_1,\dots,x_{T_n'}, \Theta \right]\right) \right)^2 \right].
  \label{eq:riskopt}
\end{equation}

\subsection{Optimisation approach}
Unfortunately, the complexity of the \gls{uks} means that evaluating any objective function involves running a full smoothing process over each training data set $\mathcal{D}_n$. For the $n^{\text{th}}$ training set, this is $\mathcal{O}(T_n)$ (since we have single dimensional latent and observation spaces) and so, for $N$ training sets, a single evaluation of an objective function is $\mathcal{O}(NT_{\text{max}})$, where $T_{\text{max}}$ is $\max_{n} T_n$.

Because of this, we use Bayesian optimisation \cite{snoek2012practical}. Bayesian optimisation uses a Gaussian process as a surrogate function to optimise low-dimensional objective functions with high evaluation cost. Since we have at most $4$ model parameters to optimise, our problem is well-suited to the Bayesian optimisation approach. The experimental setup and results for the discriminative models are detailed in Section \ref{sec:results}.

\section{Model configurations and performance results}
\begin{figure}[tb]
  \centering
  \begin{subfigure}{.32\textwidth}
    \includegraphics[width=\textwidth]{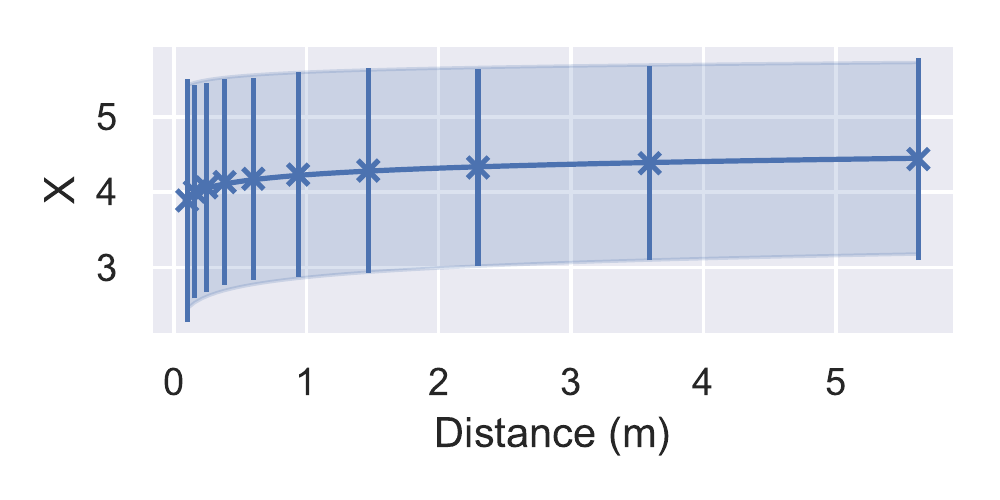} 
    \caption{Generative model.}
  \end{subfigure}
  \begin{subfigure}{.32\textwidth}
    \includegraphics[width=\textwidth]{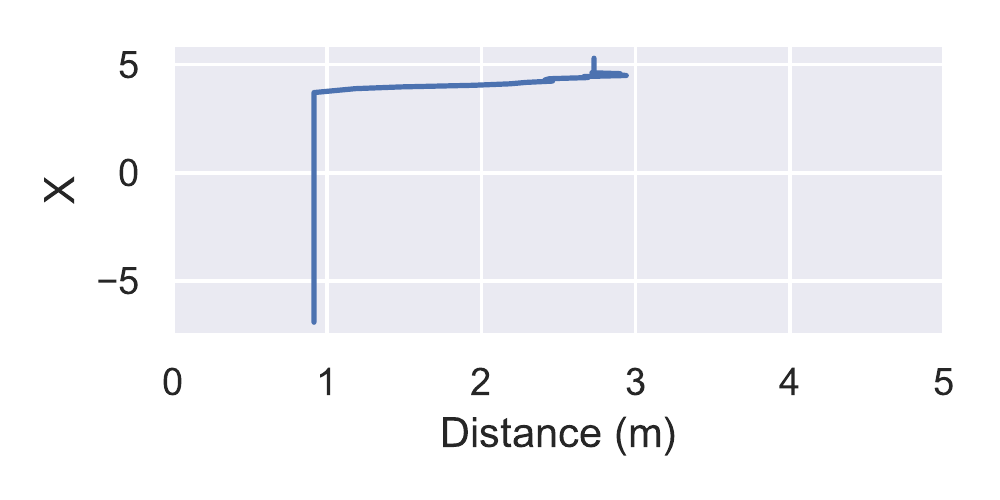} 
    \caption{GBR, MIT matrix \cite{mitmatrix}}
  \end{subfigure}
  \begin{subfigure}{.32\textwidth}
    \includegraphics[width=\textwidth]{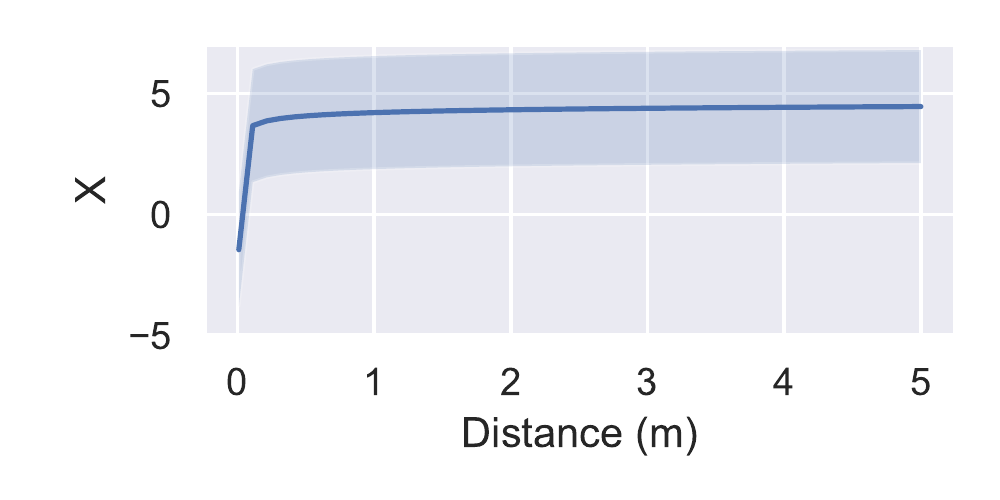} 
    \caption{Disc. model (Eq. \ref{eq:shift_model}, prox.).}
  \end{subfigure}
  \begin{subfigure}{.32\textwidth}
    \includegraphics[width=\textwidth]{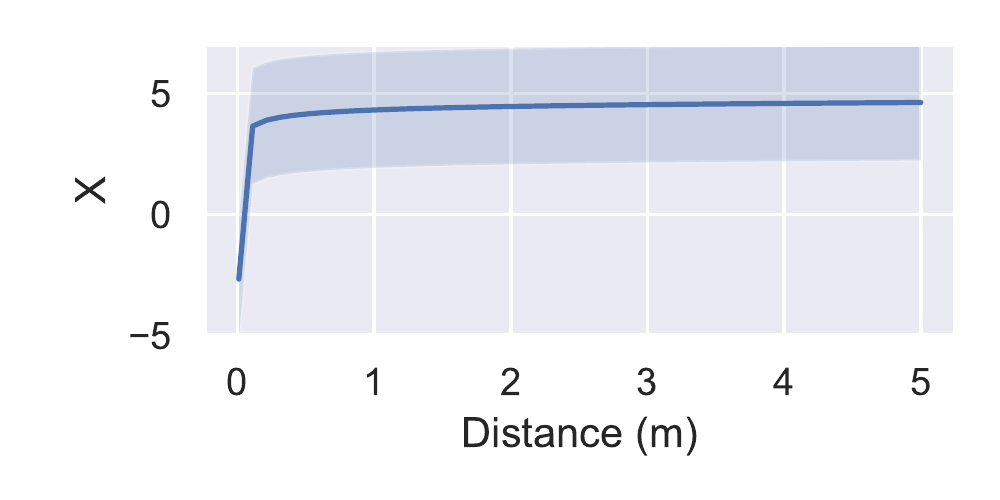} 
    \caption{Disc. model (Eq. \ref{eq:shift_model}, risk).}
  \end{subfigure}
  \begin{subfigure}{.32\textwidth}
    \includegraphics[width=\textwidth]{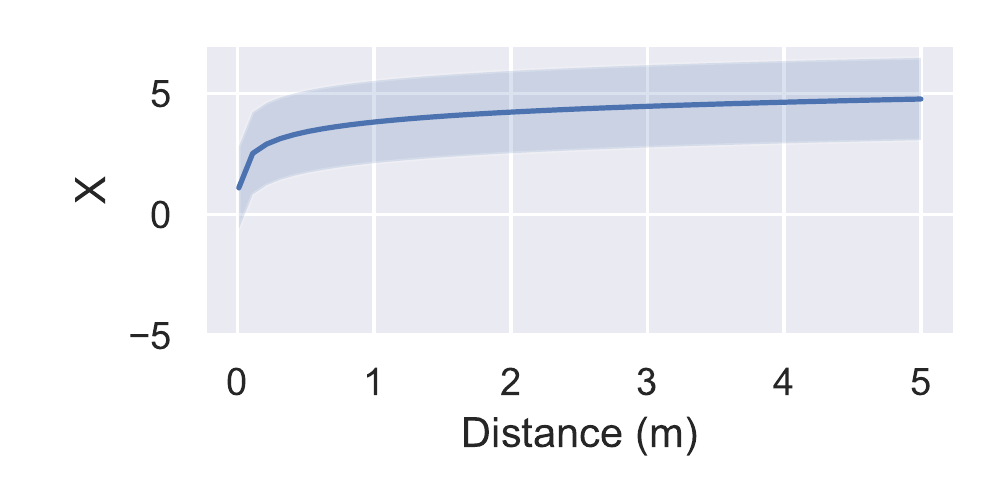} 
    \caption{Disc. model (Eq. \ref{eq:log_model}, prox.).}
  \end{subfigure}
  \begin{subfigure}{.32\textwidth}
    \includegraphics[width=\textwidth]{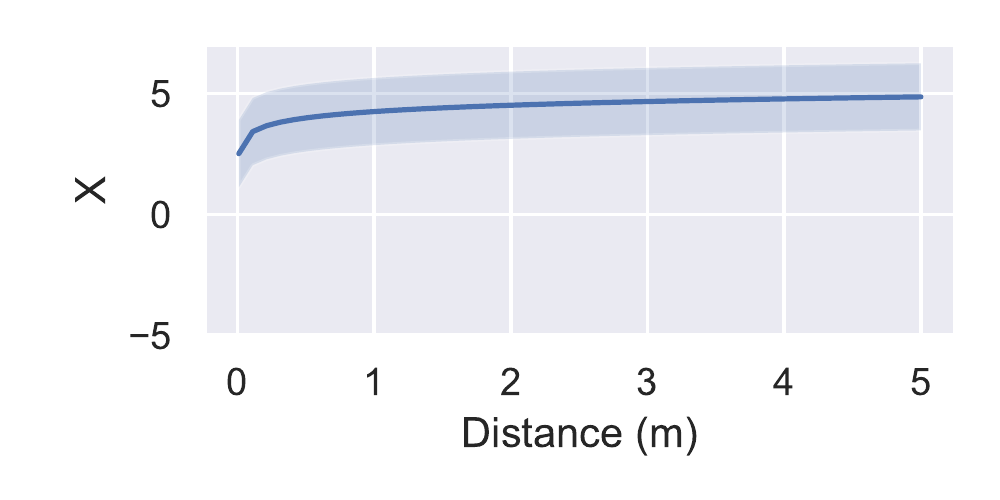} 
    \caption{Disc. model (Eq. \ref{eq:log_model}, risk).}
  \end{subfigure}
  \caption{Gaussian process data distributions for the various models. The exception is (b), which shows the gradient boosted regressor prediction of distance from \gls{rssi} (note: the axes are reversed to align with the other plots, so $d$ is a function of $X$ here). The confidence intervals mark the $0.05$ and $0.95$ quantiles of the Gaussian distributions. The generative model shows $X$ computed with $\mathbb{E}\left[ X \mid D \right]$ and $\operatorname{Var}\left( X \mid D \right)$ at a finite set of $d$ values. Interpolation is provided by standard Bayesian ridge regression on $\log(d)$. For the discriminative models, ``prox.'' means proximity optimised using Equation \ref{eq:proxopt} and ``risk'' means risk optimised using Equation \ref{eq:riskopt}.}
  \label{fig:gps}
\end{figure}
In this section, we apply both model types to various data sets. We first outline the configuration for each type and the data sets involved in parameter learning, before detailing the test data sets and presenting performance results on each for all models.

\subsection{Gradient boosted regressor}
As a benchmark for comparison, we trained a gradient boosted regressor on the MIT Matrix data set \cite{mitmatrix}. These are \gls{rssi} data captured in a variety of contexts at $8$ different distances: $3, 4, 5, 6, 8, 10, 12$ and $15$ft. There are $107$ files consisting of $118$ individual pairwise interactions.

For training, we merged all \gls{rssi} points into one set, and used $3$-fold cross validation with a test proportion of $0.33$. For the gradient boosting, we used LightGBM \cite{lightgbm} with \gls{rmse} loss function on distance, $31$ leaves and $100$ iterations. The learned prediction function is shown in Figure \ref{fig:gps} (b).

\subsection{Discriminative model configuration}
For discriminative model training data, we again used MIT's matrix data set \cite{mitmatrix}. We use $3$-fold, stratified cross-validation on the data \emph{sets} to choose the optimisation parameters for each model. The stratification is set so that at least one data set from each recorded proximity appears in both the training and validation sets. For each data set, we resample to $\Delta t = 1$s (we take the mean for multiple observations in a single time step).

For these results, we use 100 rounds of Bayesian optimisation over the full \gls{uks} from $10$ initialisation points using a Mat\'{e}rn kernel for the Gaussian process with $\nu=5/2$ and a small perturbation on the observed points ($1 \times 10^{-6}$). For the model in Equation \ref{eq:shift_model}, we used the following search ranges: $\theta_{\mu_1} \in \left[ 0.8, 1.2 \right]$; $\theta_{\mu_2} \in \left[ 0.5, 5 \right]$; $\theta_r \in \left[ 0.3, 1.5 \right]$ and $q \in \left[ 0.01, 0.05 \right]$. For the model in Equation \ref{eq:log_model}, we used: $\theta_{\mu_1} \in \left[ .01, 1 \right]$; $\theta_{\mu_2} \in \left[ 3.5, 4.5 \right]$; $\theta_r \in \left[ 0.2, 1.5 \right]$ and $q \in \left[ 0.01, 0.05 \right]$ For the optimisation process, we used the library in \cite{bo}.

For the proximity optimisation objective function in Equation \ref{eq:proxopt}, we treat each data set with equal weight, so the expectation becomes the simple mean. For the risk optimisation function in Equation \ref{eq:riskopt}, we weight each dataset according to true risk, i.e. defining
\begin{equation*}
  w_n = \sum_{t=1}^{T_n} \gamma(d_t),
\end{equation*}
where $\gamma$ is defined in Equation \ref{eq:risk}, we take the expectation in Equation \ref{eq:riskopt} over $\mathcal{D}$, with
\begin{equation*}
  p\left( \mathcal{D}_n \right) \propto w_n.
\end{equation*}

\subsection{Generative model configuration}
For the generative model, we compute $\mathbb{E}\left[ X \mid D \right]$ and $\operatorname{Var}\left( X \mid D \right)$ at a finite set of distances $\left\{ d_1,\dots,d_K \right\} \in (0, \infty)$. We fit posterior parameters $\Theta(d)$ where possible but, in the absence of empirical data, we use prior hyperparameters chosen to give appropriate uncertainty about the underlying generative processes.

\subsubsection{Environment noise $Z$}
For $Z$, we place a broad inverse-gamma prior over $\sigma_Z^2$, with $\alpha=2$ and $\beta=1/10$. This reflects our uncertainty about \gls{rssi} in any arbitrary environment. The marginal variance of $Z$ over all values for $\sigma_Z^2$ (Equation \ref{eq:varz}) gives $\operatorname{Var}\left( Z \mid \theta_Z \right) = 1/10$.

\subsubsection{Device type shifts}
\begin{figure}[tb]
  \centering
  \begin{subfigure}{.49\textwidth}
    \includegraphics[width=\textwidth]{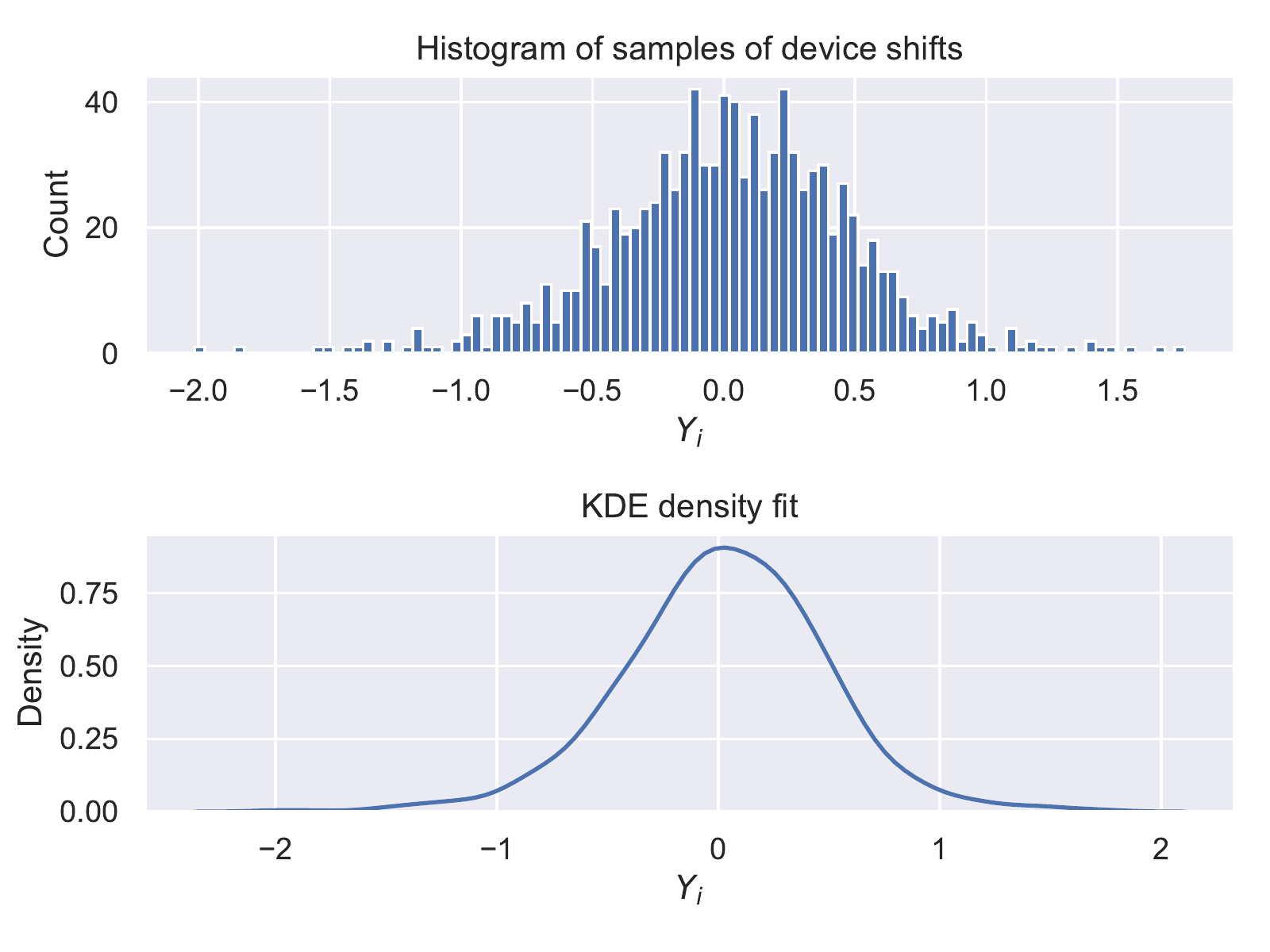}
  \end{subfigure}
  \begin{subfigure}{.49\textwidth}
    \includegraphics[width=\textwidth]{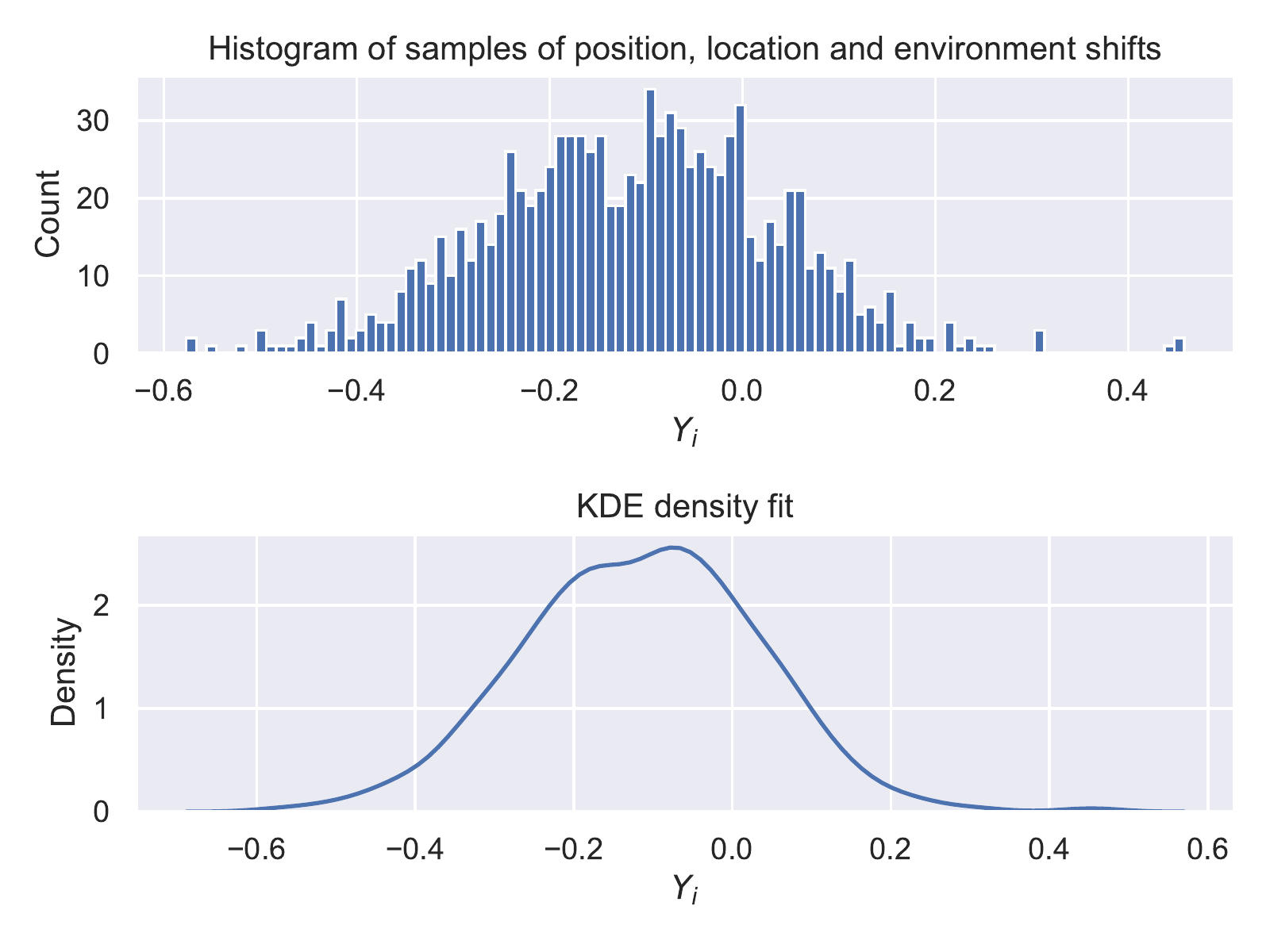}
  \end{subfigure}
  \caption{Left: $1,000$ samples from $p_{Y_i \mid d}\left( Y_i \mid \Theta \right)$ for device shifts at $d=1$m using \gls{hmc} with \gls{nuts}. Hyperparameters were set using anechoic chamber data for $729$ ($27^2$) device pairs. $\bm{\alpha}$ was set using UK mobile device market share data (see text). Right: $1,000$ samples from $p_{Y_i \mid d}\left( Y_i \mid \mathcal{D}_1,\dots,\mathcal{D}_K, \Theta \right)$ for assumed distance-invariant device position, location and environment shifts using \gls{hmc} with \gls{nuts}. Hyperparameters were set using the MIT PACT data set \cite{pact}. $\bm{\alpha}$ was set using survey data on mobile device usage (see text).}
  \label{fig:ds}
\end{figure}
We use the model in Equation \ref{eq:dshift}, with $\bm{\alpha}$ set with counts of observations of UK mobile device market share data from 2019 \cite{devicesurvey}, i.e. from the survey of $N=2,123$ respondents,
\begin{equation*}
  \bm{\alpha}_k = N(N - 1)p_rp_t,
\end{equation*}
where $p_r$ is the proportion of the $N$ respondents with device type $r$, and $p_t$ the equivalent for type $t$, with $K = r^2 = t^2$. This hyperparameter controls our belief in the specific device types in a randomly selected pair for the UK.

GSMA have provided us with calibration offset data in dBm for $27$ device makes and models. These are single observations of shifts at $d=1$m for certain transmitter/receiver device pairs in an anechoic chamber. We therefore do not have empirical data \emph{sets} for our posterior predictive model in Equation \ref{eq:dshift}, so we assume these figures set the $\mu_0$ hyperparameter for pair $k$, and that they were recorded with some uncertainty (encoded in the choices for $\lambda_0, \alpha_0, \beta_0$). Since the figures are reported in dBm, we convert these to $X$ space as follows.

We view each supplied shift $\epsilon_k$ as a $d$-invariant shift in the negative Friis transmission equation $-g(d)$. So, for $\epsilon_k > g(d)$ and $g(d) < 0$, this produces a corresponding shift $\delta_k$ in $f(d)$ as follows
\begin{align}
  \delta_k &= f(g(d) - \epsilon_k) - f(g(d)), \nonumber \\
  &= \log\left( -g(d) + \epsilon_k \right) - \log\left( -g(d) \right), \nonumber \\
  &= \log\left( 1 - \frac{\epsilon_k}{g(d)} \right),
  \label{eq:delta}
\end{align}
and we see that a constant $\epsilon_k$ results in a $\delta_k$ that varies with $d$ through $g(d)$.

We can therefore define our device type shift variable $Y_i \in \mathbb{R}$, and assume that the supplied $\delta_k$ is the $\mu_0$ parameter in Equation \ref{eq:dtmarg}. $\lambda_0, \alpha_0$ and $\beta_0$ encode our uncertainty about $Y_i$, and we set them to $\lambda_0=1, \alpha_0=2$ and $\beta_0=1/10$ to give small variance about $\mu_0$. Samples from $p_{Y_i \mid d}\left( Y_i \mid \Theta(d) \right)$ for $d=1$m are shown in Figure \ref{fig:ds}.

\subsubsection{Antenna gain shifts}
In addition to the calibration data supplied by GSMA, an estimated gain figure $\epsilon$ (in dBm) for the calibration reference device was also supplied. We follow the approach of the previous section to convert this to $X$ space,
\begin{equation}
  \delta = \log\left( 1 - \frac{\epsilon}{g(d)} \right),
  \label{eq:delta1}
\end{equation}
and set the $d$-specific hyperparameter $\mu_0 = \delta$. We again choose $\alpha_0=2, \beta_0=1/10$ and $\lambda_0=1$.

\subsubsection{Device position, location and environment shifts}
We use the model in Equation \ref{eq:mixt}, with the following hyperparameter settings. We assume the environment factor is split into two: indoors and outdoors. We set the indoors' probability to $p(E = \text{indoors})=0.869$ and the outdoors' probability to the complement, which are taken from the National Human Activity Survey (NHAPS) \cite{klepeis2001national}.

For device location, we use the data from \cite{rescuetime} and assume a mobile device is not concealed for $8$ hours (sleep) plus $3.25$ hours in active use plus $8$ hours not in use but nearby, e.g. working; leaving $4.75$ hours with the device being concealed in a pocket or bag. Thus, we set the concealed probability to $p(L = \text{concealed}) = 4.75/24$, and the not concealed probability to the complement.

For device position, we assume equal belief to all orientation angles in a 2D plane. So, for any finite $K$-partitioning of $[0, 2\pi)$ into intervals $I_1,\dots,I_K$, we set $p(P \in I_k) = \lvert I_k \rvert/2\pi$. 

With these, the $k^{\text{th}}$ component of the Dirichlet hyperparameter becomes, for  realisation $\left( p, l, e \right)_k$,
\begin{equation*}
  \bm{\alpha}_k = Np\left( e \right)p\left( l \right)p\left( e \right),
\end{equation*}
where $N$ is a pseudocount, which we set to $10$.

For the remaining posterior predictive hyperparameters, we use the PACT datasets provided by MIT \cite{pact}. These contain measurements for two ``real world'' environment classes (1 outdoor and 3 indoor, with the 3 indoor sets merged) over two transmit location classes (3 concealed and 1 in hand, with the 3 concealed sets merged) over 8 angles of orientation. We use these data to set the parameters in Equation \ref{eq:ppparams} -- with $\mu_0 = 0, \lambda_0=\beta_0=1/10$ and $\alpha_0=2$ -- as follows.

We use the reference data sets recorded over location (concealed and in-hand) and positions (8 angles) in an anechoic chamber. We assume these are observations of noisy reference \gls{rssi}, and fit a normal distribution to $X := \log(-R)$,
\begin{equation}
  X_{l, p} \sim \mathcal{N}(\hat{\mu}_{l, p}, \hat{\sigma}_{l, p}^2), 
  \label{eq:envnorm}
\end{equation}
where the parameters are the sample mean and (unbiased) variance for each of the 16 reference sets. For the 32 other data sets, we observe shift variables as follows. For a given data set $\mathcal{D}_{p,l,e} = \left\{ x_1,\dots,x_{N_{p,l,e}} \right\}$, we draw
\begin{equation*}
  \hat{x}_1,\dots,\hat{x}_{N_{e,l,p}} \sim \mathcal{N}(\hat{\mu}_{l, p}, \hat{\sigma}_{l, p}^2), 
\end{equation*}
and use the observed shifts $y_i = x_i - \hat{x}_i$. These data are then used to compute the parameters for the $t$-distributions using Equation \ref{eq:ppparams}. See Figure \ref{fig:ds} for samples from the full distribution of $Y_i$ using these posterior parameters, and the mixing weights' distribution parameters described above.

\subsubsection{Approximating $\mathbb{E}\left[ X \mid D \right]$ and $\operatorname{Var}\left( X \mid D \right)$}
\begin{figure}[tb]
  \centering
  \includegraphics[width=.7\textwidth]{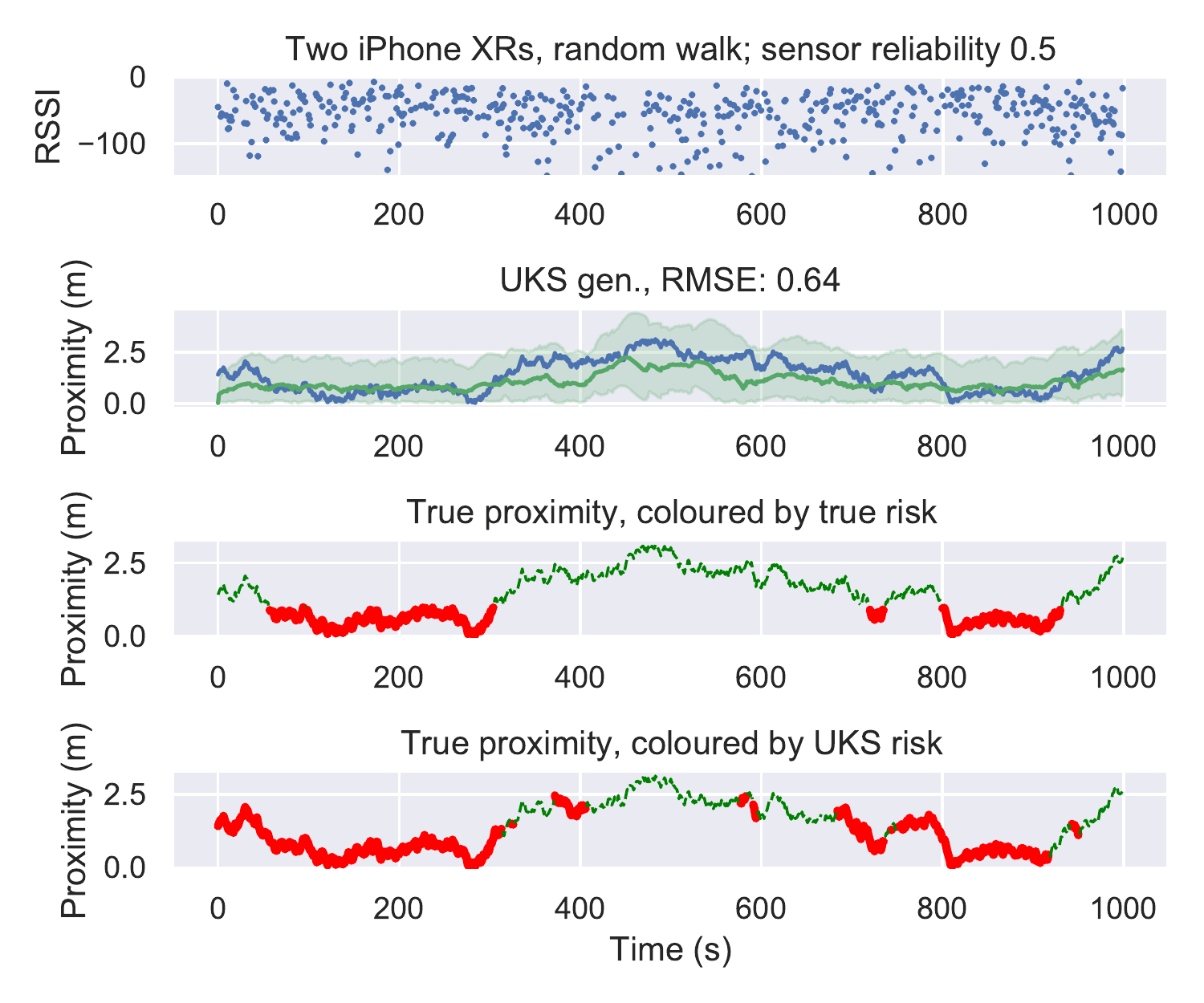}
  \caption{The \gls{uks} with generative observation model and $q=.09$ applied to simulated random walk data. Here, two iPhone XR devices undertake a random walk at on a circle with radius $2$m for $1,000$ seconds. We fit a sampling model with iPhone XR device types known (that is, without the mixture component over device type) from the MIT H0H1 data sets \cite{h0h1}. This results in a Gaussian process with $\mu(d) = 0.21\log(d) + 3.92$ and $\sigma^2 = 0.33$. \gls{rssi} samples are then generated at each time step from $\operatorname{log-normal}(\mu(d), \sigma^2)$. In this example, half the observations are removed randomly to simulate imperfect sensor reliability. The topmost plot shows the \gls{rssi} data; the second plot shows the \gls{uks} with moment-matched gamma distribution $0.05$ and $0.95$ quantiles; the third and fourth plots show true and inferred risk coloration respectively -- high risk, i.e. when within $1$m of each other, is the thicker, solid red line; low risk is the thinner, dashed green line. Note the imputation of the \gls{uks} where there are missing observations.}
  \label{fig:random_walk}
\end{figure}

With the distributions of $Y_i \mid \mathcal{D}_1,\dots,\mathcal{D}_K, \Theta$ and the value of $\operatorname{Var}\left( Z \mid \theta_Z \right)$ set in the previous sections, we can approximate the expectation and variance of $X$ with Equations \ref{eq:exp} and \ref{eq:var}. For this we use use \gls{hmc} with \gls{nuts} to estimate the expectations in Equations \ref{eq:exp} and \ref{eq:var}. The resulting Gaussian process under the computed estimates is shown in Figure \ref{fig:gps} (a). 

Figure \ref{fig:random_walk} shows the \gls{uks} with generative observation model tracking proximity from noisy simulated data using \gls{rssi} generated from devices in the MIT H0H1 data set \cite{h0h1}.

\subsection{Test data sets}
For performance evaluation, we use the MIT H0H1 data set \cite{h0h1} and the Trinity College Dublin data sets from \cite{leith2020coronavirus}.

\subsubsection{MIT H0H1 data set}
This data set consists of \gls{rssi} captures from $26$ ``high risk'' scenarios (H1), and $19$ ``low risk'' scenarios (H0). We define a scenario to be an interaction between a device pair, and some of the raw data files contain multiple device interactions. There are iPhone and Android devices present in the data sets.

In the H1 scenario, participants were asked to stay within $6$ft of each other for $15$ minutes. In the H0 scenario, they were asked to stay at least $10$ft apart for 15 min. In each scenario, participants were instructed to interact with each other normally in multiple environments, including: outdoors, indoors and sat at a table. Participants were also allowed to use their mobile phones as normal throughout the study.

We do not know if participants genuinely strayed over the instructed boundaries, nor do we know if each raw data file was intended to capture a single interaction. We include all mobile phone interactions in all files regardless.

Since we have proximity and risk bounds only, we cannot measure exact inference, but we know that the models should infer proximity $\le 6$ft for H1, and $\ge 10$ft for H0, and thus ``high risk'' and ``low risk'' respectively.

\begin{figure}[tb]
  \centering
  \begin{subfigure}{.50\textwidth}
    \includegraphics[width=\textwidth]{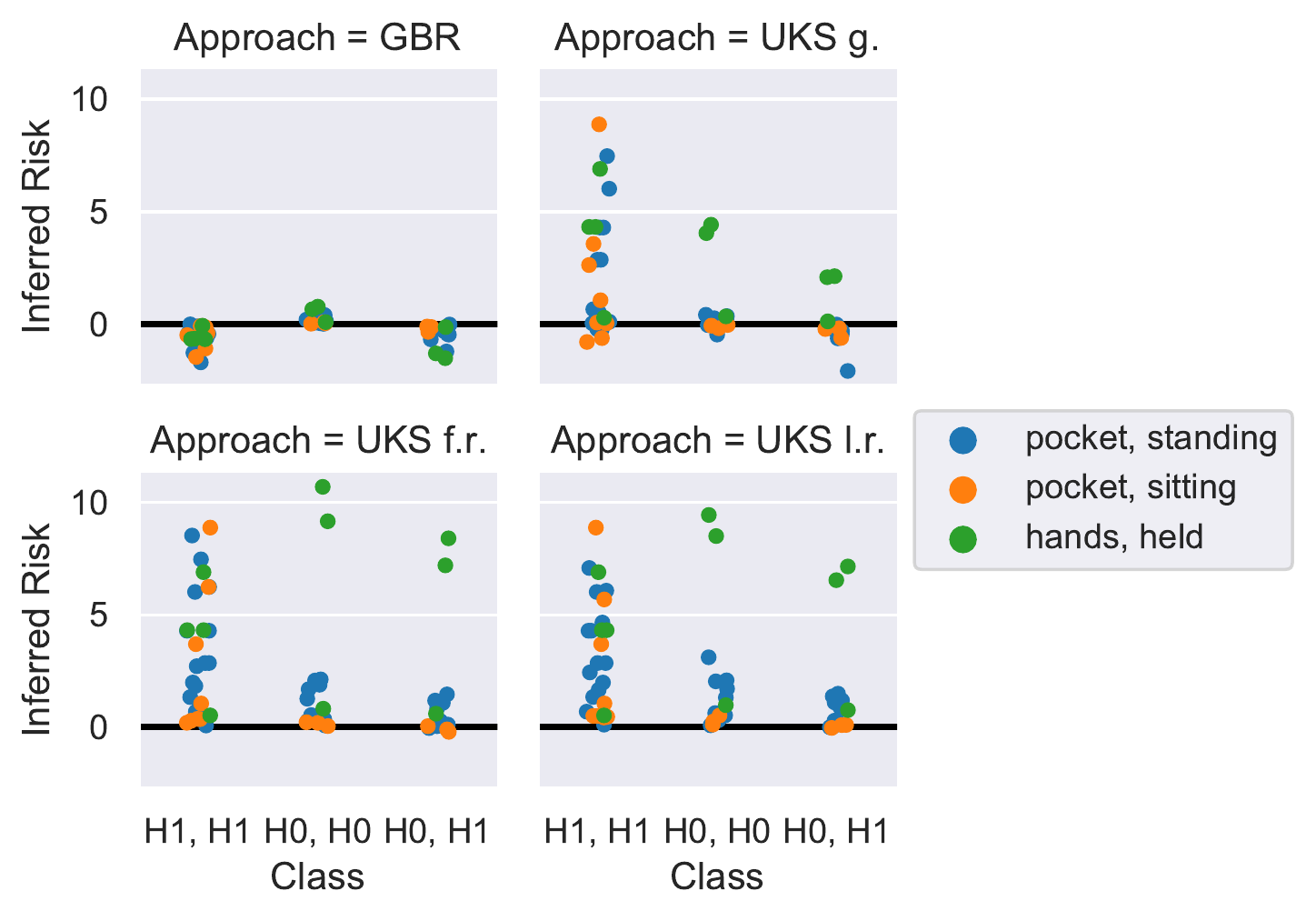} 
  \end{subfigure}
  \begin{subfigure}{.48\textwidth}
    \includegraphics[width=\textwidth]{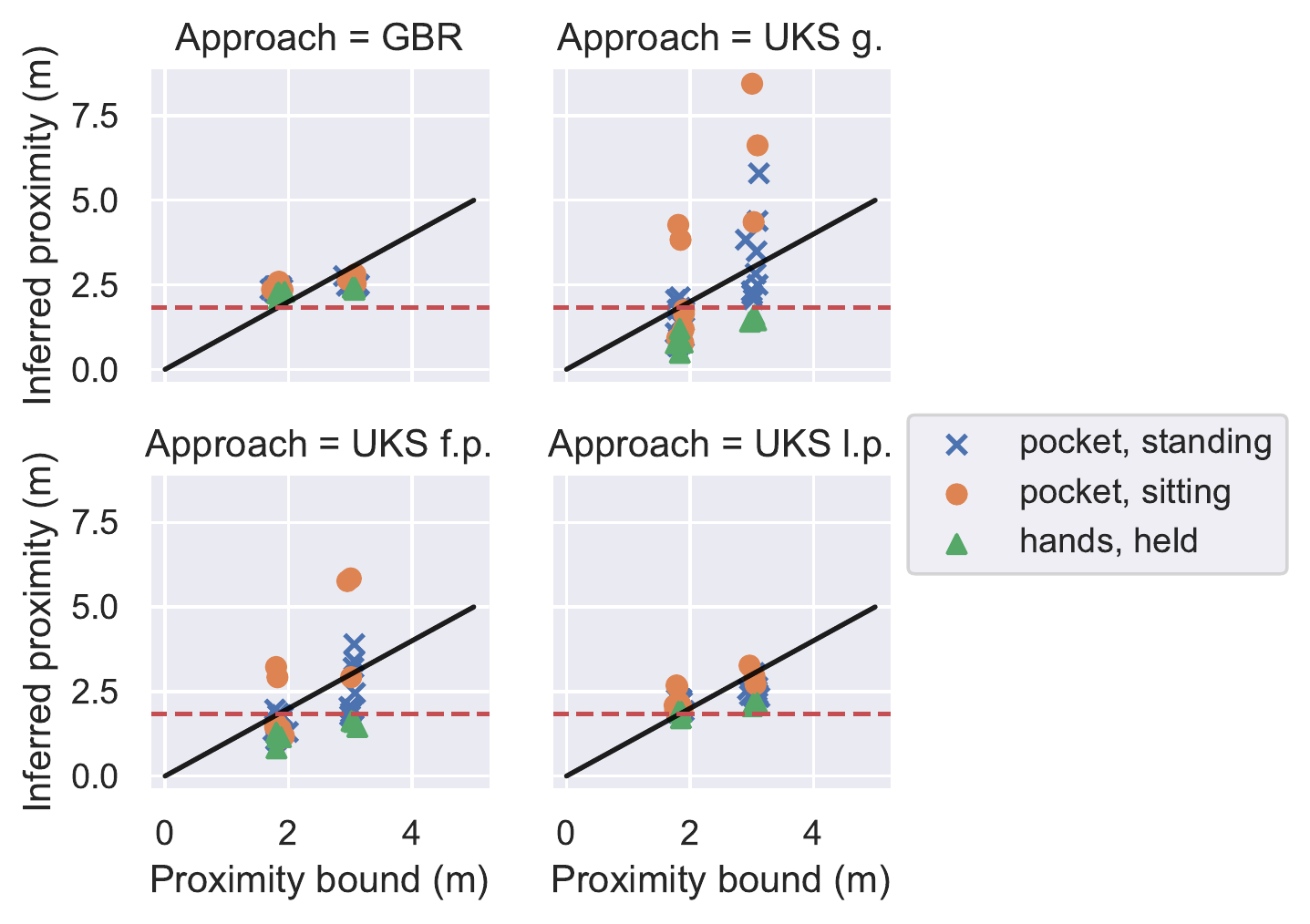} 
  \end{subfigure}
  \caption{Left: relative risk, i.e. inferred risk minus true risk (bound), for MIT H0H1. H1, H1 means the high-risk scenario with high-risk threshold. H0, H0 is the low-risk scenario with low-risk threshold. H0, H1 is the low-risk scenario with high-risk threshold. Right: inferred proximity against true proximity (bound) for MIT H0H1. The two columns of points (with jitter) are the true bounds for H1 and H0 respectively. The red dashed line is the H1 proximity bound.  GBR is the gradient boosted regressor; \gls{uks} g. is the generative model; \gls{uks} f.r./f.p. are the discriminative models in Equation \ref{eq:shift_model} optimised for risk/proximity. \gls{uks} l.r./l.p. are the equivalent for Equation \ref{eq:log_model}. See text for further details on plot interpretation.}
  \label{fig:h0h1}
\end{figure}

\subsubsection{Trinity College data set}
This data set consists of \gls{rssi} readings in a number of settings, some of which are laboratory settings and some real-world settings. We use the real-world, or scenario settings, which consist of $14$ sets of \gls{rssi} data in environments such as supermarkets, desks, public transport and walking in public. An approximate ground truth proximity is labelled for each set, though we do not know if participants rigidly adhered to this proximity throughout the capture. There are only Google Pixel 2 devices present in the data sets.

\subsection{Performance results}
\label{sec:results}
\begin{figure}[tb]
  \centering
  \begin{subfigure}{.32\textwidth}
    \includegraphics[width=\textwidth]{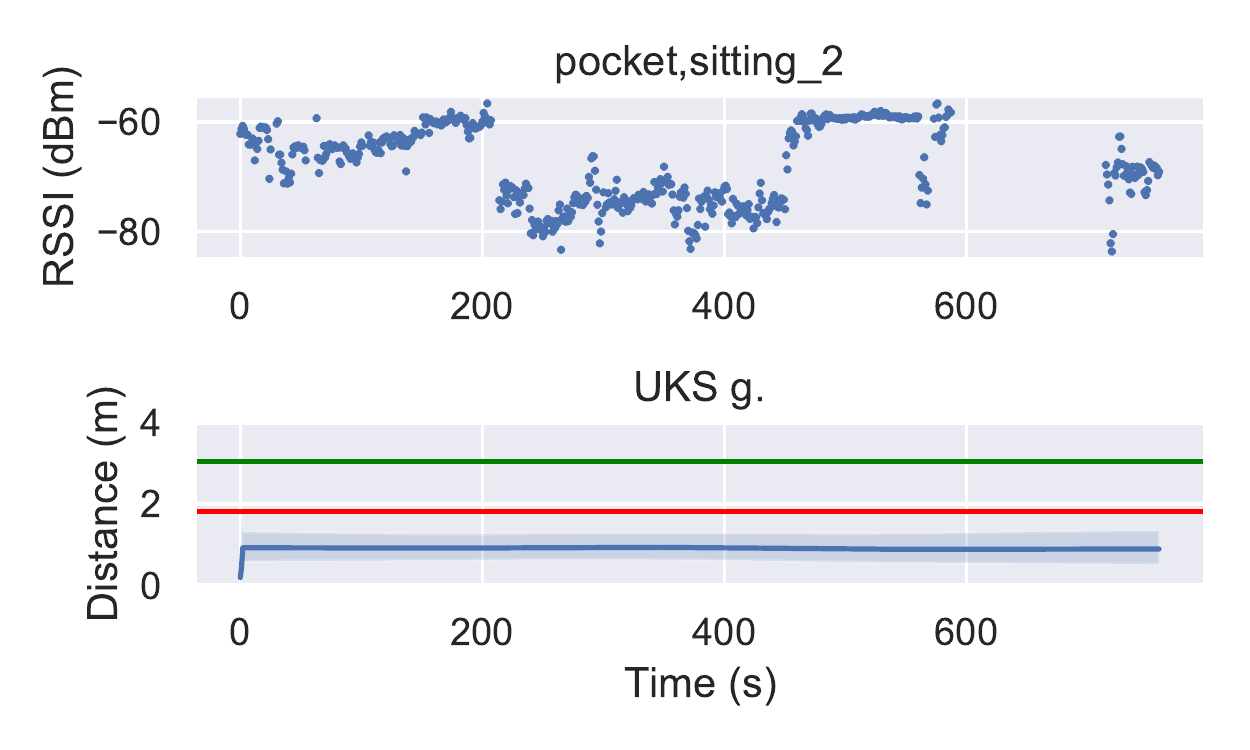} 
  \end{subfigure}
  \begin{subfigure}{.32\textwidth}
    \includegraphics[width=\textwidth]{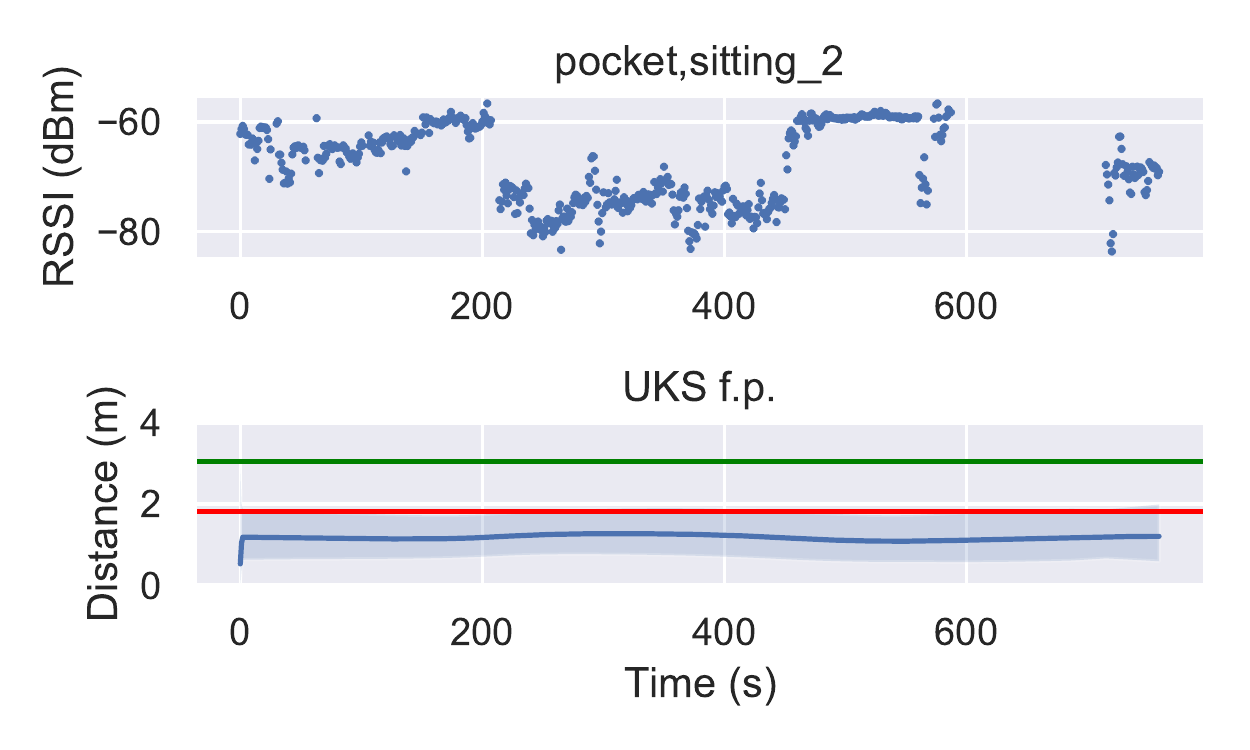} 
  \end{subfigure}
  \begin{subfigure}{.32\textwidth}
    \includegraphics[width=\textwidth]{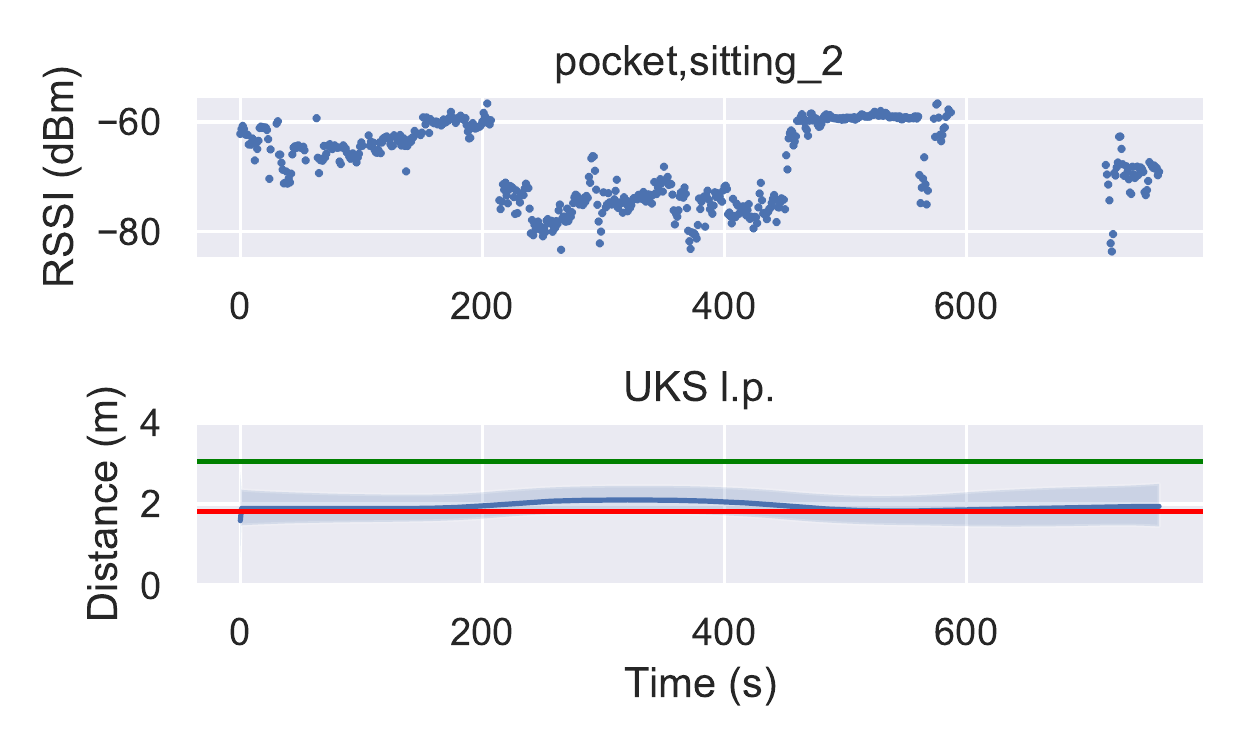} 
  \end{subfigure}
  \begin{subfigure}{.32\textwidth}
    \includegraphics[width=\textwidth]{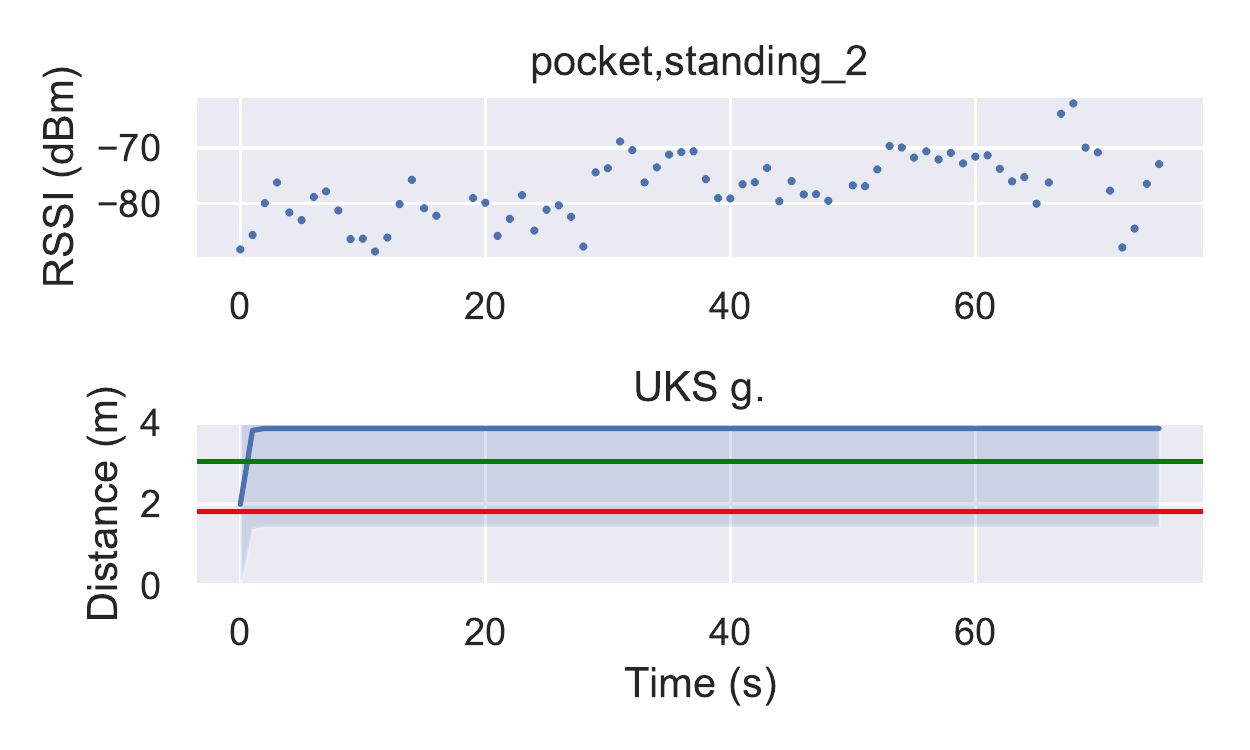} 
  \end{subfigure}
  \begin{subfigure}{.32\textwidth}
    \includegraphics[width=\textwidth]{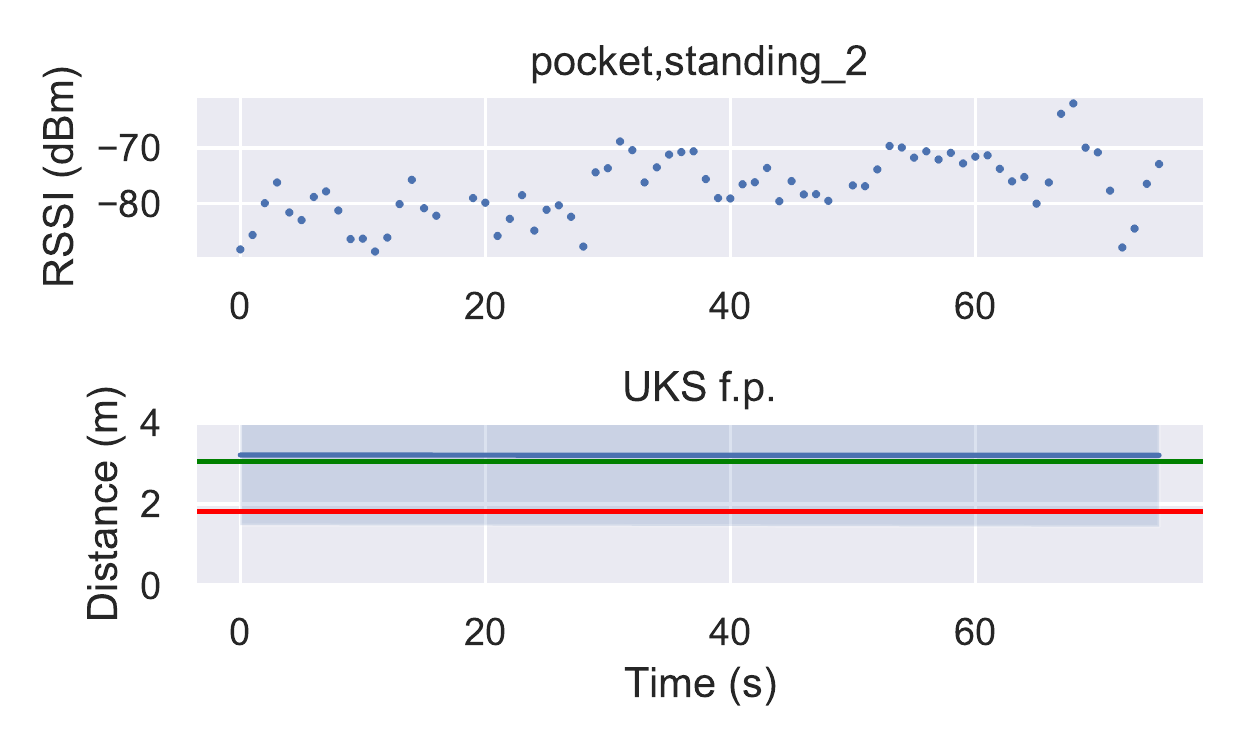} 
  \end{subfigure}
  \begin{subfigure}{.32\textwidth}
    \includegraphics[width=\textwidth]{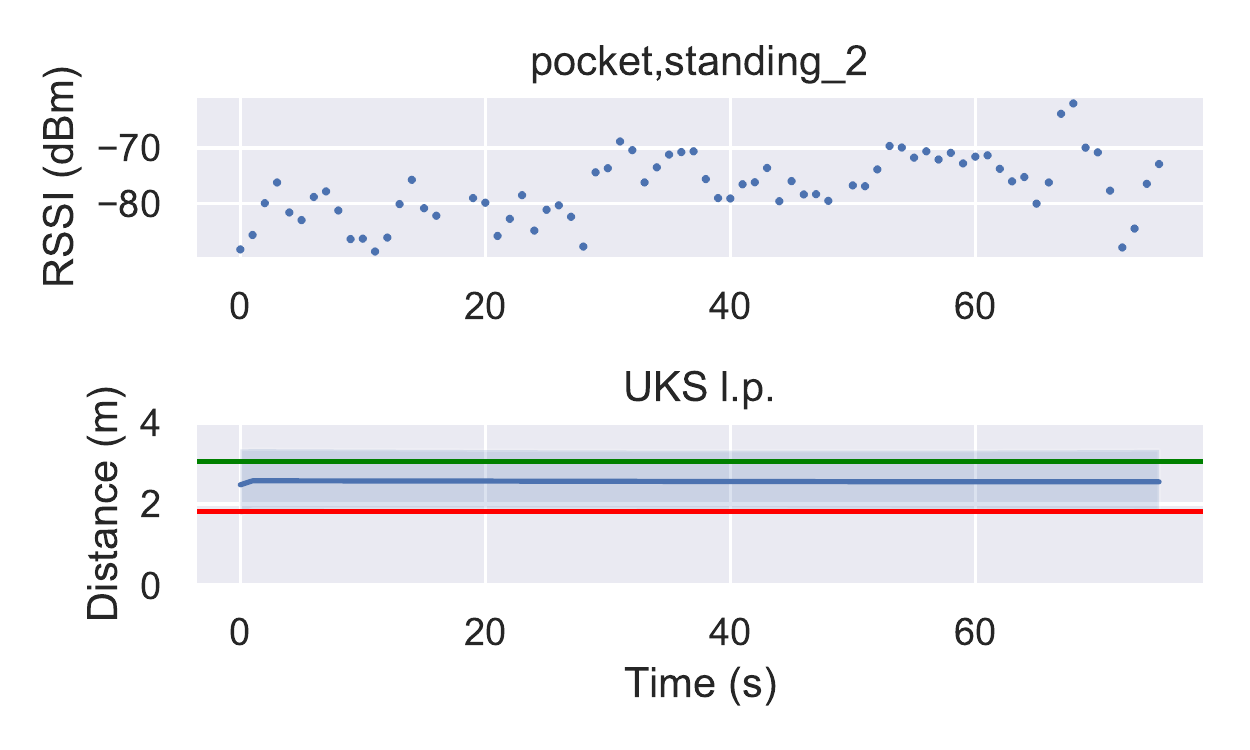} 
  \end{subfigure}
  \caption{Time series of observed \gls{rssi} and \gls{uks} output (mean with $0.05$ and $0.95$ quantiles of a moment-matched gamma distribution) on one H1 example and one H0 example from MIT H0H1. Top row: H1 (high risk scenario); bottom row: H0 (low risk scenario); first column: \gls{uks} with generative model; second column: \gls{uks} with discriminative model (Equation \ref{eq:shift_model}); third column: \gls{uks} with discriminative model (Equation \ref{eq:log_model}). The red horizontal line is the H1 threshold $6$ft, and the green horizontal line is the H0 threshold.}
  \label{fig:h0h1_series}
\end{figure}
\begin{figure}[tb]
  \centering
  \includegraphics[width=\textwidth]{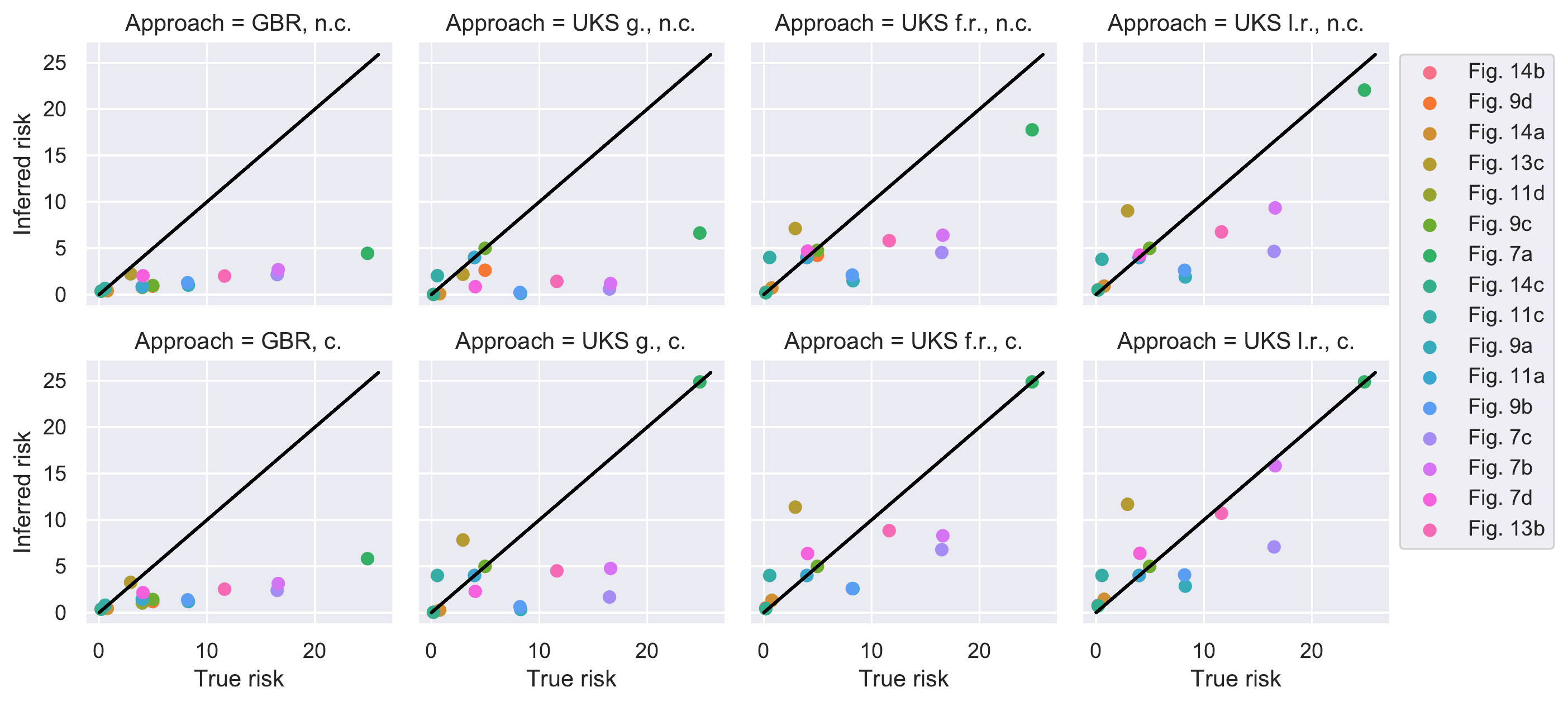}
  \caption{Trinity College data, with inferred risk against true risk. The top row shows the results for raw \gls{rssi} data. The bottom row shows the results when the \gls{rssi} are corrected with the knowledge of mobile device types (Google Pixel 2). GBR is the gradient boosted regressor; \gls{uks} g. is the generative model; \gls{uks} f.r./f.p. are the discriminative models in Equation \ref{eq:shift_model} optimised for risk/proximity. \gls{uks} l.r./l.p. are the equivalent for Equation \ref{eq:log_model}. The plot legend refers to figures in \cite{leith2020coronavirus}. The n.c. and c. refer to ``not corrected'' and ``corrected'' respectively.}
  \label{fig:trinity}
\end{figure}

Figure \ref{fig:h0h1} shows the results for the MIT H0H1 data. Each point is a single scenario, coloured/marked by description. For the risk plots, we use relative risk, i.e. inferred risk minus the true risk bound. Since we do not know the true proximities -- only their bounds -- we can assess performance by visualising where the models place the scenarios above or below the $0$ line.

For high risk, i.e. H1, points should be placed on or above the line, with increasing error the further below the line. For H0, points should be placed on or below the line, with increasing error the further above the line. A good proximity classifier would put all points above the black line for H1, H1 and all points below the line for H0, H0. A good risk classifier would put all points in H0, H1 below the line.

The \gls{roc} \gls{auc} for the approaches are: gradient boosting regressor: 0.5; \gls{uks} g.: 0.823; \gls{uks} f.p.: 0.756; \gls{uks} f.r.: 0.6; \gls{uks} l.p.: 0.538; and \gls{uks} l.r.: 0.567.

For the proximity plot, a good proximity classifier would put all points in the H1 column below the black x-y line, and all the points in the H0 column above the black x-y line. A good risk classifier would put all points in the H0 column above the red dashed line (the H1 threshold).

Figure \ref{fig:h0h1_series} shows example time series for the \gls{uks} with different models on an H1 and H0 scenario. Figure \ref{fig:trinity} shows inferred risk against true risk for the Trinity College Dublin scenarios in \cite{leith2020coronavirus}, the figures from which correspond to the plot legend labels in Figure \ref{fig:trinity}.

\section{Discussion}
\label{sec:discussion}
Here we discuss the implications and limitations of the results in the previous section. The key finding is that good prediction of proximity and risk can be achieved by treating \gls{rssi} sequences and using posterior inference of proximity $D_t$ given the entire sequence of observations $x_1,\dots,x_{T'}$ rather than $x_t$ alone. By using a \gls{uks}, we can undertake this inference with nonlinear observation models; in this case Gaussian processes, which also encode uncertainty that propagates through to the posterior distribution over $D_t$. Given the single dimensions of both state space and observations, inference for a periodic sequence $D_1,\dots,D_T$ can be achieved in linear time, i.e. $\mathcal{O}(T)$.

Using sequential modelling with Gaussian process data distributions outperforms simpler thresholding approaches such as the Exposure Notification API (cf. the effectively threshold-based gradient boosting \gls{auc} of 0.5 on MIT H0H1, which seems to corroborate the findings in \cite{leith2020tram}).

We next we compare the performance and suitability of proximity inference vs direct risk inference, before analysing the results of the generative and discriminative approaches. Next we acknowledge the implications of making a log-normal assumption for the distribution of $-R$, before discussing general limitations and potential areas for improvement in future work.

\subsection{Proximity vs risk}
Our chief intention is to infer posterior proximity given observed \gls{rssi} values, but there is an argument that predicting infection risk directly is more pertinent, especially given the application to the Covid-19 pandemic. The plots for the MIT H0H1 data in Figure \ref{fig:h0h1} and the Trinity College data in Figure \ref{fig:trinity} show how the duration component of risk can make some encounters significantly more important to classify correctly. This implies that jointly inferring proximity and duration is arguably more important than proximity alone since, for example, a long duration at a farther proximity can equate to a shorter duration at closer proximity, and a classifier that seeks to predict close encounters well at the cost of incorrectly predicting farther ones may not achieve the desired effect of good overall infection risk prediction. We have not inferred duration here beyond the time duration of the scenarios in the test data, but an area for further work would be to better improve duration inference from real world \gls{rssi} observations.

\subsection{Generative vs discriminative models}
The results show that using the \gls{uks} with either a generative or discriminative model will likely outperform simple classification approaches, but there is a question as to which model is more appropriate. The generative model is the best performing approach for the MIT H0H1 data (Figure \ref{fig:h0h1}), but the discriminative models outperform the generative model in the Trinity College data (Figure \ref{fig:trinity}). There is an argument that the generative model is more general, since the discriminative models are limited by the training data, but the hyperparameters of the generative model are also computed from example data. There is arguably more flexibility to the generative model, since arbitrary numbers of shift variables $Y_i$ can be added, but there is no strong evidence in our results to recommend choosing one over the other.

The question of which discriminative model to use is also not answered definitively, but the results in Figures \ref{fig:h0h1} and \ref{fig:trinity} show marginally better performance using the form of Equation \ref{eq:shift_model} over Equation \ref{eq:log_model}. There are also limitations introduced by the search restrictions of Bayesian optimisation, and there may be better parameters for these models that were not found in the optimisation process.

It is perhaps unsurprising that the scenarios with the greatest error are where the mobile device is in the pocket with the individual sitting (Figure \ref{fig:h0h1}; false negatives), and where the device is in the hand (Figure \ref{fig:h0h1}; false positives).

\subsection{Log-normal $-R$ vs Gaussian $R$}
The main results in the paper assume that $-R$ is a log-normal random variable, but there is a valid argument that $R$ should be normally distributed, e.g. in the log-distance path loss model for radio propagation. Our justification for using the log-normal distribution was based on empirical evidence of a long tail in observed real-world \gls{rssi} values, plus the assumption that transmission power should be at most $0$dBm for \gls{ble} in mobile devices.

We have replicated all results using a normally distributed $R$, i.e. $X := R$, and these can be seen in Appendix \ref{sec:appendix}. There is a small difference in performance, with the log-normal model performing slightly better in general on the test data sets. The notable exception is the generative model on the Trinity College data (Figure \ref{fig:trinity} vs Figure \ref{fig:trinity_gaussian}), but the average performance of the log-normal appears to be slightly better. We conjecture that this is due to the resilience against \gls{rssi} fluctuations due to the long tail of the log-normal data distribution (compare the steadiness of the inferred proximity in Figures \ref{fig:h0h1_series} and \ref{fig:h0h1_series_gaussian}).

\subsection{Limitations and potential improvements}
Since we have not attempted to infer duration here, an obvious next step would be to focus on this; perhaps by attempting to partition \gls{rssi} data into sessions. We have also not considered other machine learning classifiers beyond a gradient boosting regressor, and it is entirely possible that a well-trained neural network could perform well, though we argue that much of the performance stems from the sequential modelling, and a sequential neural network may be a better choice. (These approaches are of course limited by access to good quality training data.) The other advantage of the \gls{uks} is uncertainty quantification, since we have posterior probability distributions over $D_t$ and can report our confidence in the inference given the many sources of uncertainty in the data and underlying dynamics.

Other potential improvements could include: acquiring more, high quality training data for the models; exploring more complex \gls{uks} approaches, though it may be prudent to keep state space dimensions low since inference is na\"ively $\mathcal{O}(Td^3)$; optimising discriminative model parameters using approaches other than Bayesian optimisation; exploring data distribution forms other than log-normal (and Gaussian), e.g. the compound $k$ distribution; and analysing performance as \gls{rssi} quality deteriorates, either due to noise or by intention for conservation of power.

The inference of proximity from \gls{ble} \gls{rssi} is a difficult problem, and more learning and validation data sets captured in varied scenarios (including from simulations) can only benefit any modelling approach.

\section{Conclusion}
In this paper, we presented a novel approach to inferring proximity from \gls{ble} \gls{rssi} using a \gls{uks} with Gaussian process data distribution. This is especially relevant to mobile phone applications designed to tackle the Covid-19 pandemic, which rely on good inference of infection risk; itself a function of proximity. We outlined two approaches to characterising the data distribution: a generative model, which directly computes sources of variability in observations; and a discriminative model, which optimises model parameters on example training data. There is no strong evidence to choose the generative approach over the discriminative one (or \emph{vice versa}). Risk and proximity inference performance on two real-world data sets -- MIT H0H1 and Trinity College Dublin -- show that the \gls{uks} outperforms a more traditional gradient boosted regressor model. Our work to date offers an insight into well established mechanisms for probabilistic modelling of one of the key latent factors, that of proximity. We recognise that this needs to be considered in the wider context of health policy, ethical and other technical considerations, when responsibly deploying novel technology of this kind.

\bibliographystyle{abbrv}
\bibliography{refs}
\newpage
\appendix
\section{Results using a Gaussian distribution on $R$ directly}
There is some debate about a suitable distribution for $X$. In the paper, we assumed a log-normal distribution on raw \gls{rssi} $R$ and used a log transform to define the normally distributed $X := \log(-R)$. This was motivated by asymmetric forms observed in empirical data, with long tails going to $-\infty$ and -- under the assumption of at most a $0$dBm transmission power -- support on $(-\infty, 0]$.

In this supplement, we replicate all the results in the paper under a direct Gaussian model on $R$, i.e., we define $X := R$. This shows that the log-normal model is more robust to noise, but that performance on the test data sets is comparable.

\section{Data distribution form}
The base function $f$ in Equation \ref{eq:base_function} is now equivalent to Equation \ref{eq:friis}, i.e. 
\begin{equation*}
  f(d) = g(d).
\end{equation*}

\section{Model configurations and performance results}
\label{sec:appendix}
In this section, we present the model configurations and results under the direct Gaussian observation model.
\begin{figure}[tb]
  \centering
  \begin{subfigure}{.32\textwidth}
    \includegraphics[width=\textwidth]{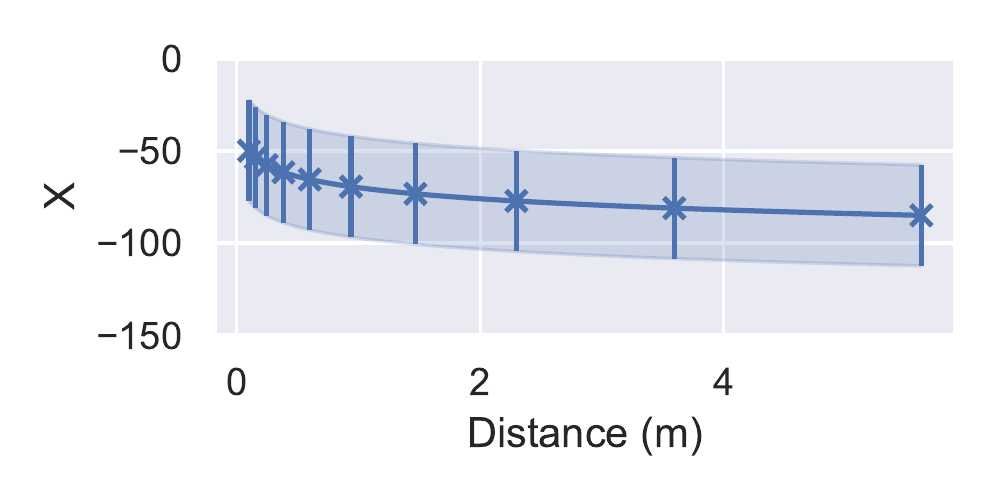} 
    \caption{Generative model.}
  \end{subfigure}
  \begin{subfigure}{.32\textwidth}
    \includegraphics[width=\textwidth]{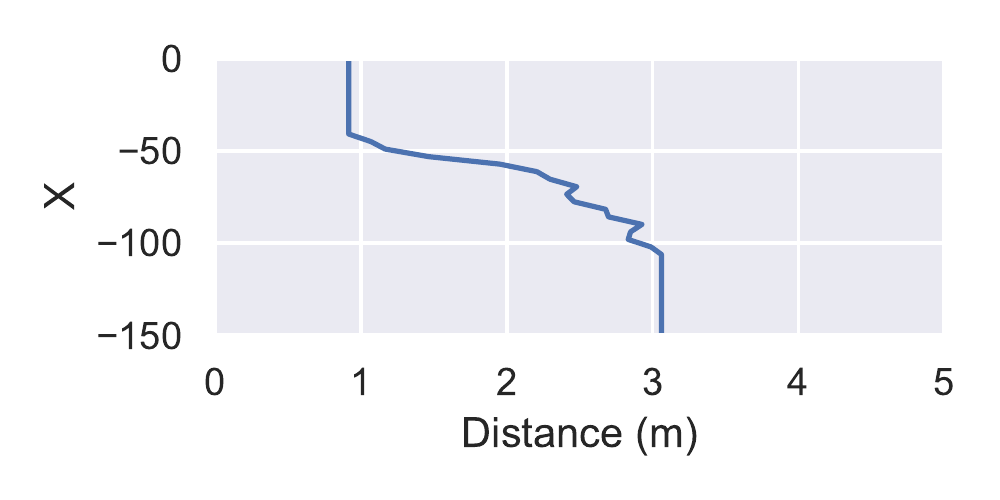} 
    \caption{GBR, MIT matrix \cite{mitmatrix}}
  \end{subfigure}
  \begin{subfigure}{.32\textwidth}
    \includegraphics[width=\textwidth]{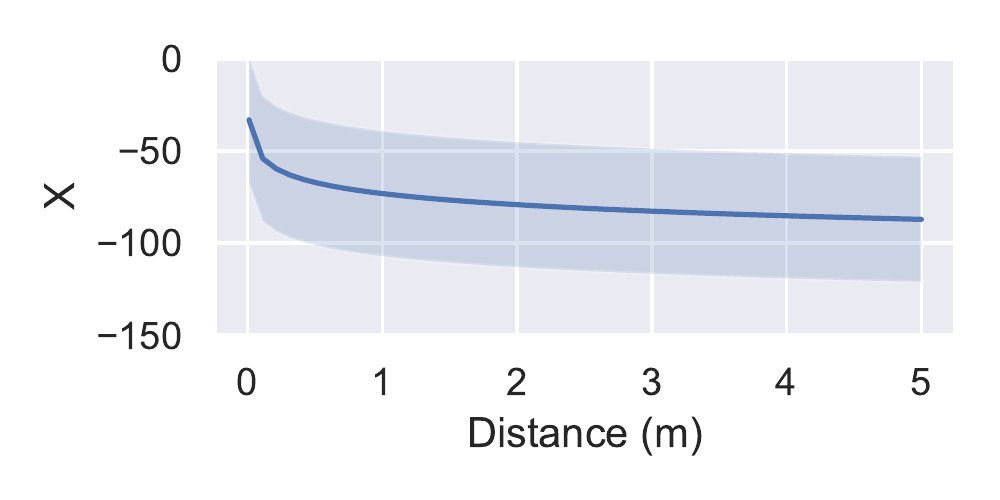} 
    \caption{Disc. model (Eq. \ref{eq:shift_model}, prox.).}
  \end{subfigure}
  \begin{subfigure}{.32\textwidth}
    \includegraphics[width=\textwidth]{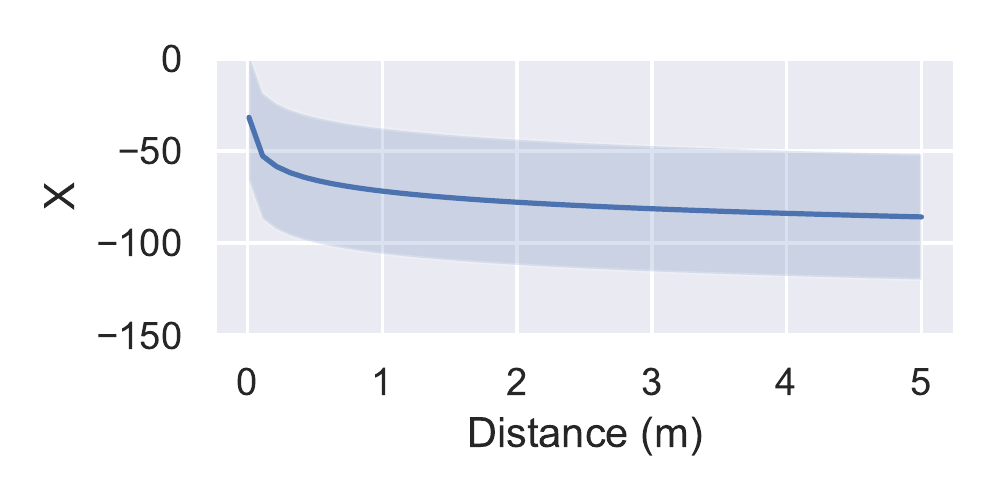} 
    \caption{Disc. model (Eq. \ref{eq:shift_model}, risk).}
  \end{subfigure}
  \begin{subfigure}{.32\textwidth}
    \includegraphics[width=\textwidth]{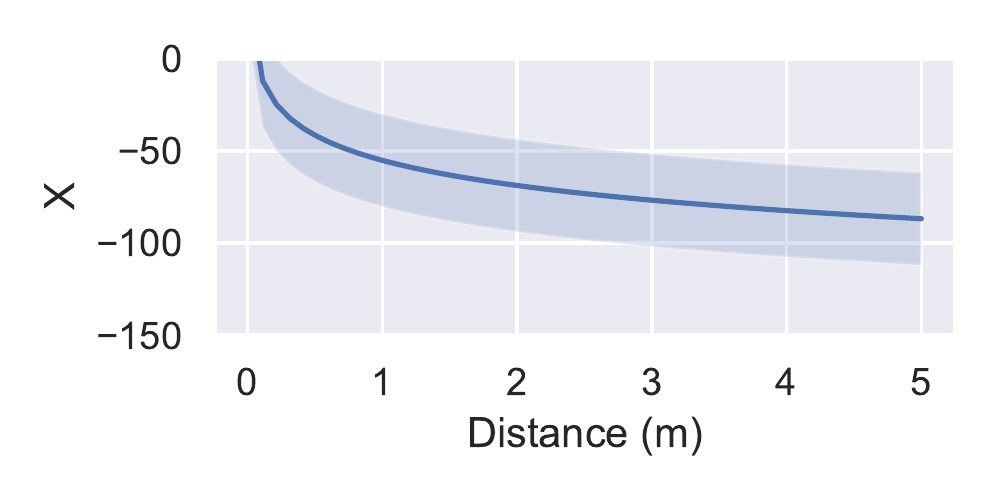} 
    \caption{Disc. model (Eq. \ref{eq:log_model}, prox.).}
  \end{subfigure}
  \begin{subfigure}{.32\textwidth}
    \includegraphics[width=\textwidth]{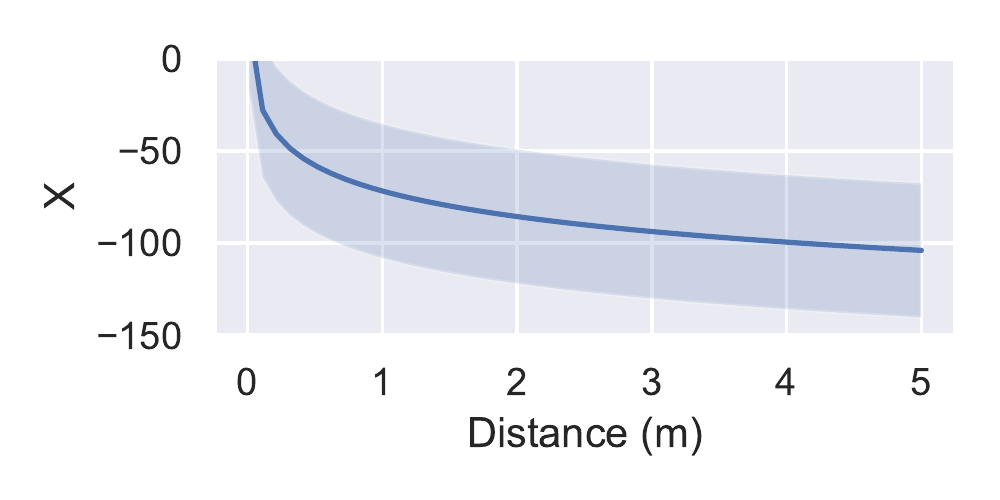} 
    \caption{Disc. model (Eq. \ref{eq:log_model}, risk).}
  \end{subfigure}
  \caption{Gaussian process data distributions for the various models. The exception is (b), which shows the gradient boosted regressor prediction of distance from \gls{rssi} (note: the axes are reversed to align with the other plots, so $d$ is a function of $X$ here). The confidence intervals mark the $0.05$ and $0.95$ quantiles of the Gaussian distributions. The generative model shows $X$ computed with $\mathbb{E}\left[ X \mid D \right]$ and $\operatorname{Var}\left( X \mid D \right)$ at a finite set of $d$ values. Interpolation is provided by standard Bayesian ridge regression on $\log(d)$. For the discriminative models, ``prox.'' means proximity optimised using Equation \ref{eq:proxopt} and ``risk'' means risk optimised using Equation \ref{eq:riskopt}.}
  \label{fig:gps_gaussian}
\end{figure}

\subsection{Discriminative model configuration}
For these results, we use 100 rounds of Bayesian optimisation over the full \gls{uks} from $10$ initialisation points using a Mat\'{e}rn kernel for the Gaussian process with $\nu=5/2$ and a small perturbation on the observed points ($1 \times 10^{-6}$). For the model in Equation \ref{eq:shift_model}, we used the following search ranges: $\theta_{\mu_1} \in \left[ 1., 1. \right]$; $\theta_{\mu_2} \in \left[ -100, -10 \right]$; $\theta_r \in \left[ 0, 300 \right]$ and $q \in \left[ 0.01, 0.05 \right]$. For the model in Equation \ref{eq:log_model}, we used: $\theta_{\mu_1} \in \left[ -20, -1 \right]$; $\theta_{\mu_2} \in \left[ -100, -10 \right]$; $\theta_r \in \left[ 0, 300 \right]$ and $q \in \left[ 0.01, 0.05 \right]$ For the optimisation process, we used the library in \cite{bo}.

\subsection{Generative model configuration}
The $\delta_k$ shifts in Equations \ref{eq:delta} and \ref{eq:delta1} simply map directly to negative $\epsilon_k$, i.e.
\begin{equation*}
  \delta_k = -\epsilon_k.
\end{equation*}

The normal distribution in Equation \ref{eq:envnorm} is fit to $X := R$ rather than $X := \log(-R)$. The variance in Equation \ref{eq:varz} is set to 10dB, i.e. $\beta = 10, \alpha = 2$.
\begin{figure}[tb]
  \centering
  \begin{subfigure}{.49\textwidth}
    \includegraphics[width=\textwidth]{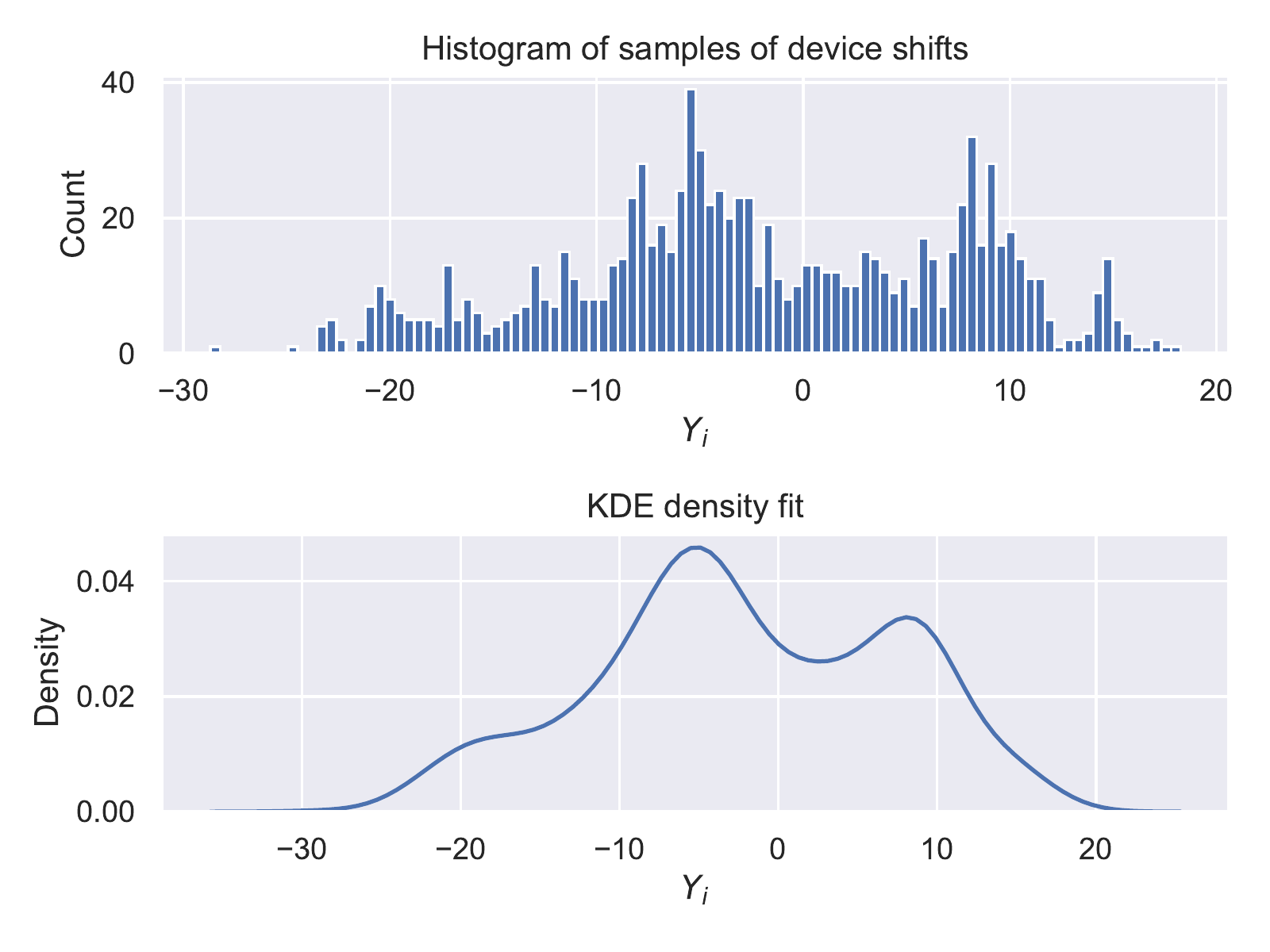}
  \end{subfigure}
  \begin{subfigure}{.49\textwidth}
    \includegraphics[width=\textwidth]{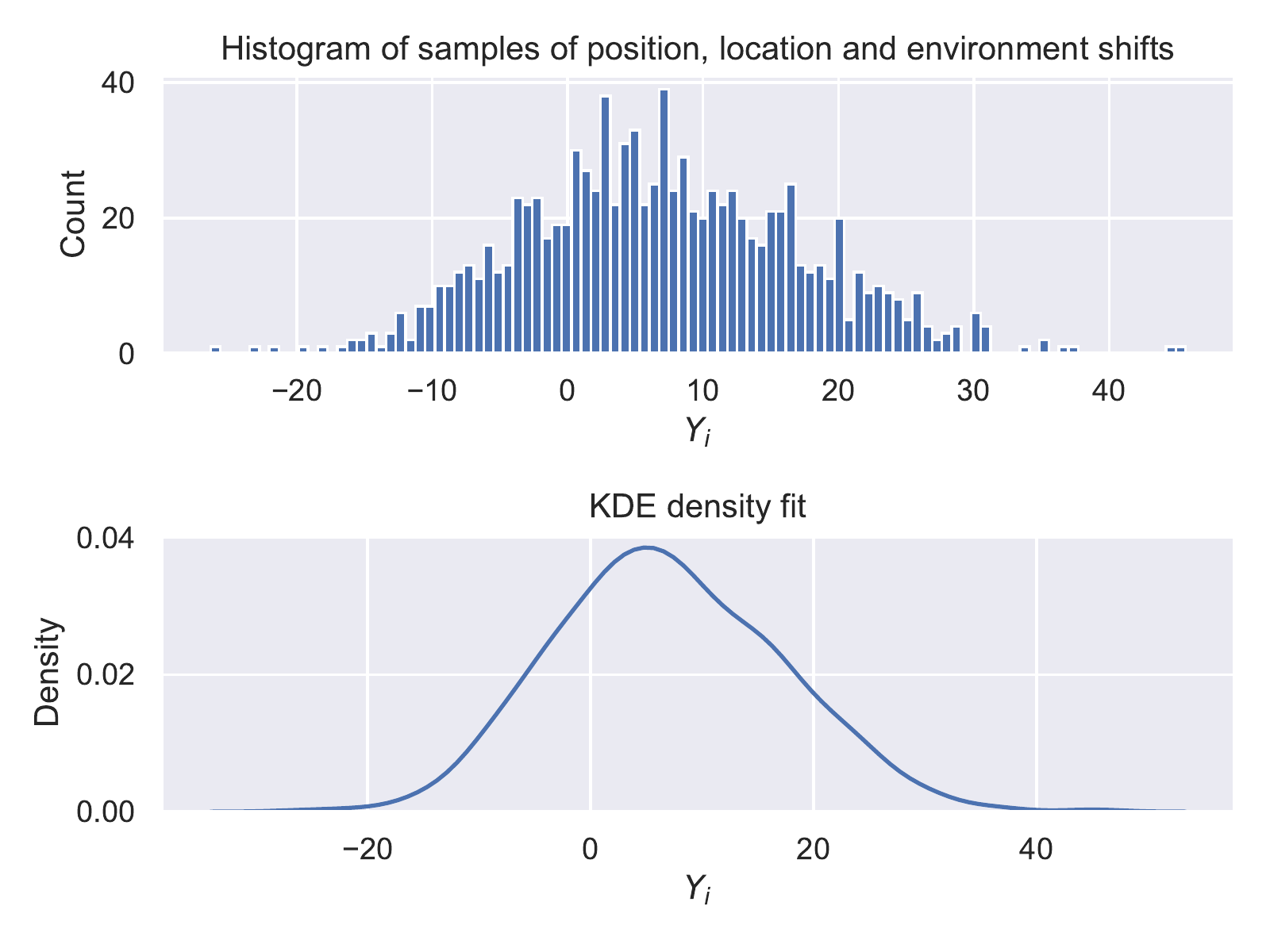}
  \end{subfigure}
  \caption{Left: $1,000$ samples from $p_{Y_i \mid d}\left( Y_i \mid \Theta \right)$ for device shifts at $d=1$m using \gls{hmc} with \gls{nuts}. Hyperparameters were set using anechoic chamber data for $729$ ($27^2$) device pairs. $\bm{\alpha}$ was set using UK mobile device market share data (see text). Right: $1,000$ samples from $p_{Y_i \mid d}\left( Y_i \mid \mathcal{D}_1,\dots,\mathcal{D}_K, \Theta \right)$ for assumed distance-invariant device position, location and environment shifts using \gls{hmc} with \gls{nuts}. Hyperparameters were set using the MIT PACT data set \cite{pact}. $\bm{\alpha}$ was set using survey data on mobile device usage (see text).}
  \label{fig:ds_gaussian}
\end{figure}

\subsection{Performance results}
\begin{figure}[tb]
  \centering
  \includegraphics[width=.7\textwidth]{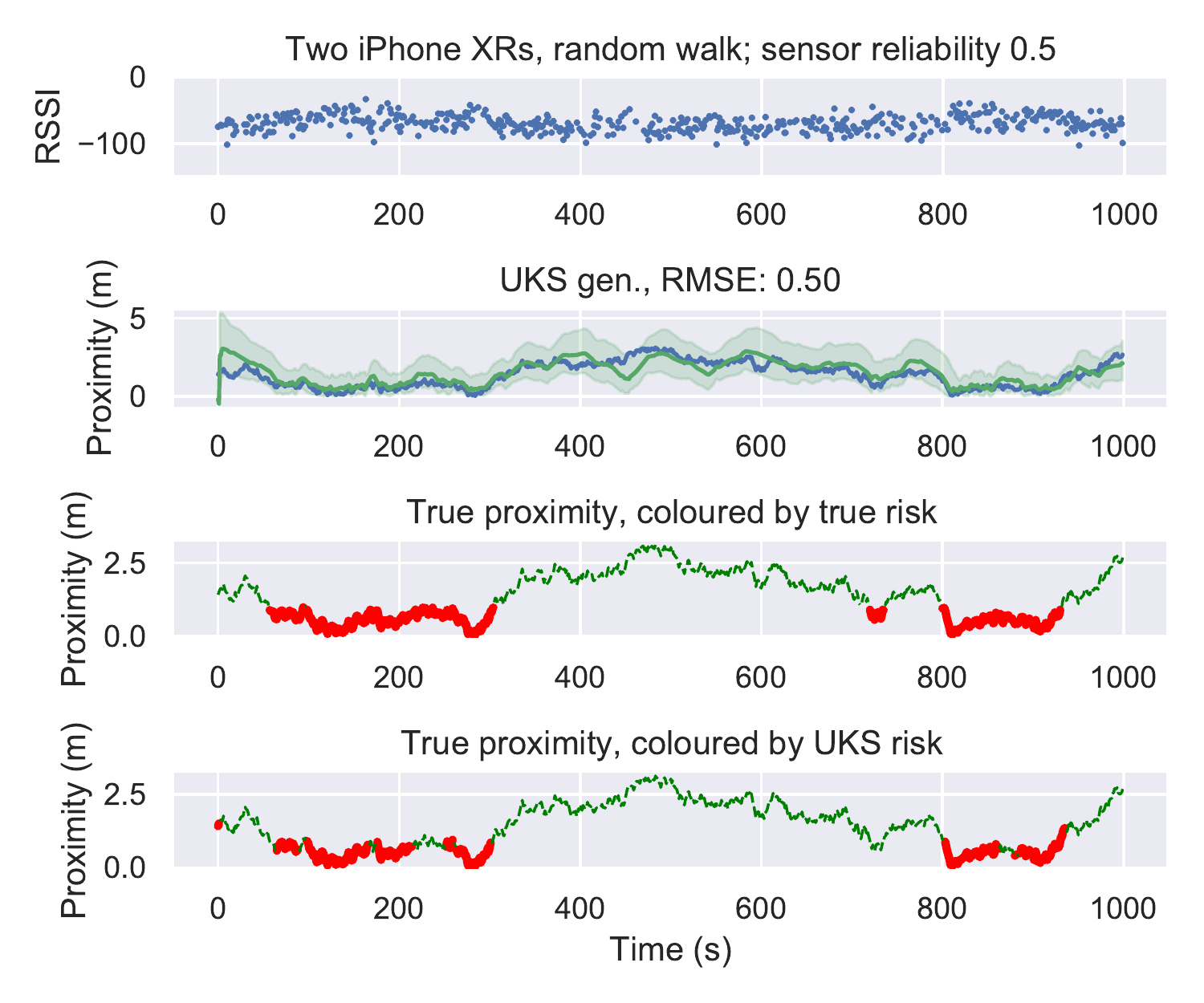}
  \caption{The \gls{uks} with generative observation model and $q=.09$ applied to simulated random walk data. Here, two iPhone XR devices undertake a random walk on a circle with radius $2$m for $1,000$ seconds. We fit a sampling model with iPhone XR device types known (that is, without the mixture component over device type) from the MIT H0H1 data sets \cite{h0h1}. This results in a Gaussian process with $\mu(d) = -8.69\log(d) - 67.9 $ and $\sigma^2 = 97.03$. \gls{rssi} samples are then generated at each time step from $\mathcal{N}(\mu(d), \sigma^2)$. In this example, half the observations are removed randomly to simulate imperfect sensor reliability. The topmost plot shows the \gls{rssi} data; the second plot shows the \gls{uks} with moment-matched gamma distribution $0.05$ and $0.95$ quantiles; the third and fourth plots show true and inferred risk respectively -- high risk, i.e. when within $1$m of each other, is the thicker, solid red line; low risk is the thinner, dashed green line. Note the imputation of the \gls{uks} where there are missing observations.}
  \label{fig:random_walk_gaussian}
\end{figure}

\begin{figure}[tb]
  \centering
  \begin{subfigure}{.50\textwidth}
    \includegraphics[width=\textwidth]{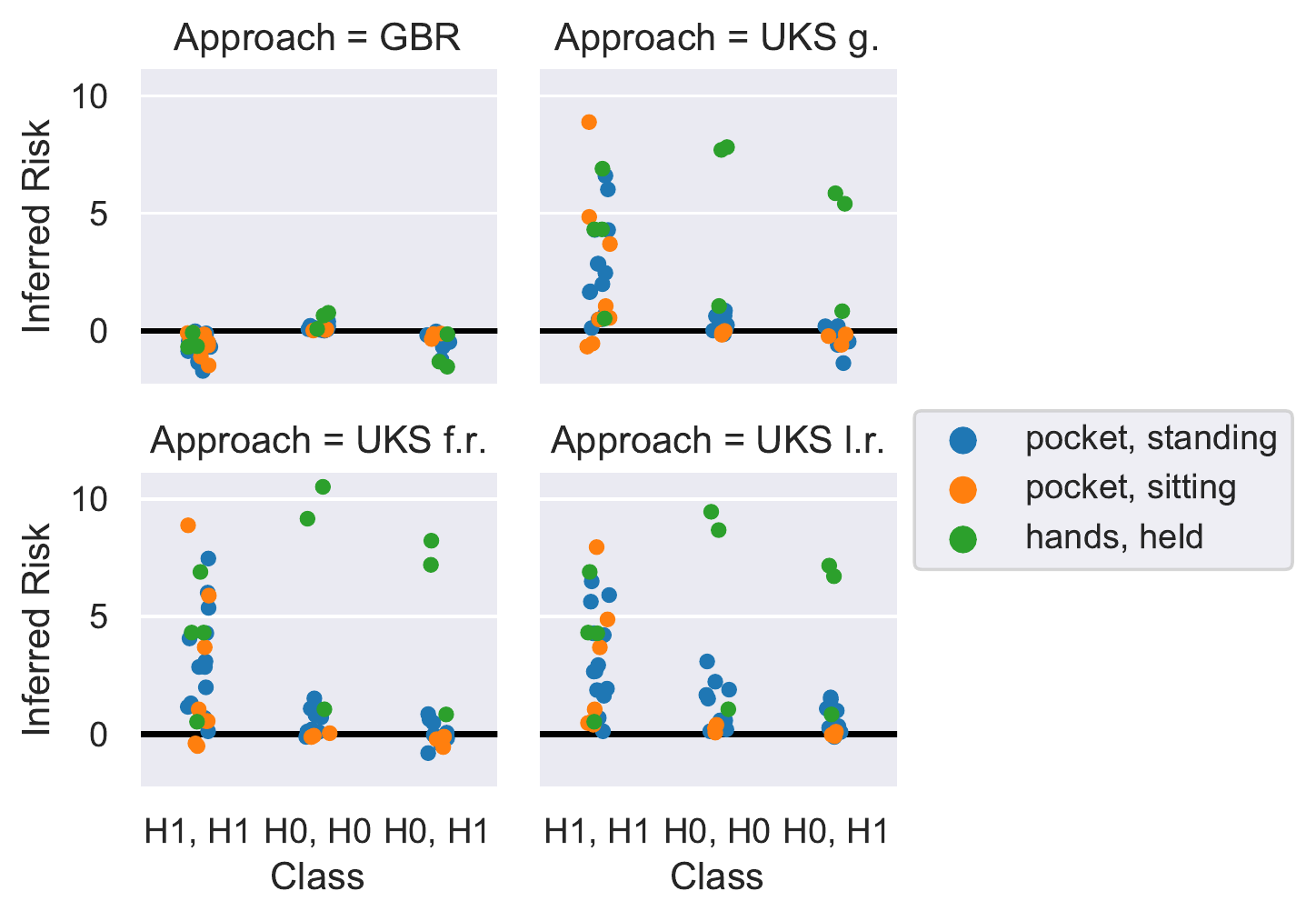} 
  \end{subfigure}
  \begin{subfigure}{.48\textwidth}
    \includegraphics[width=\textwidth]{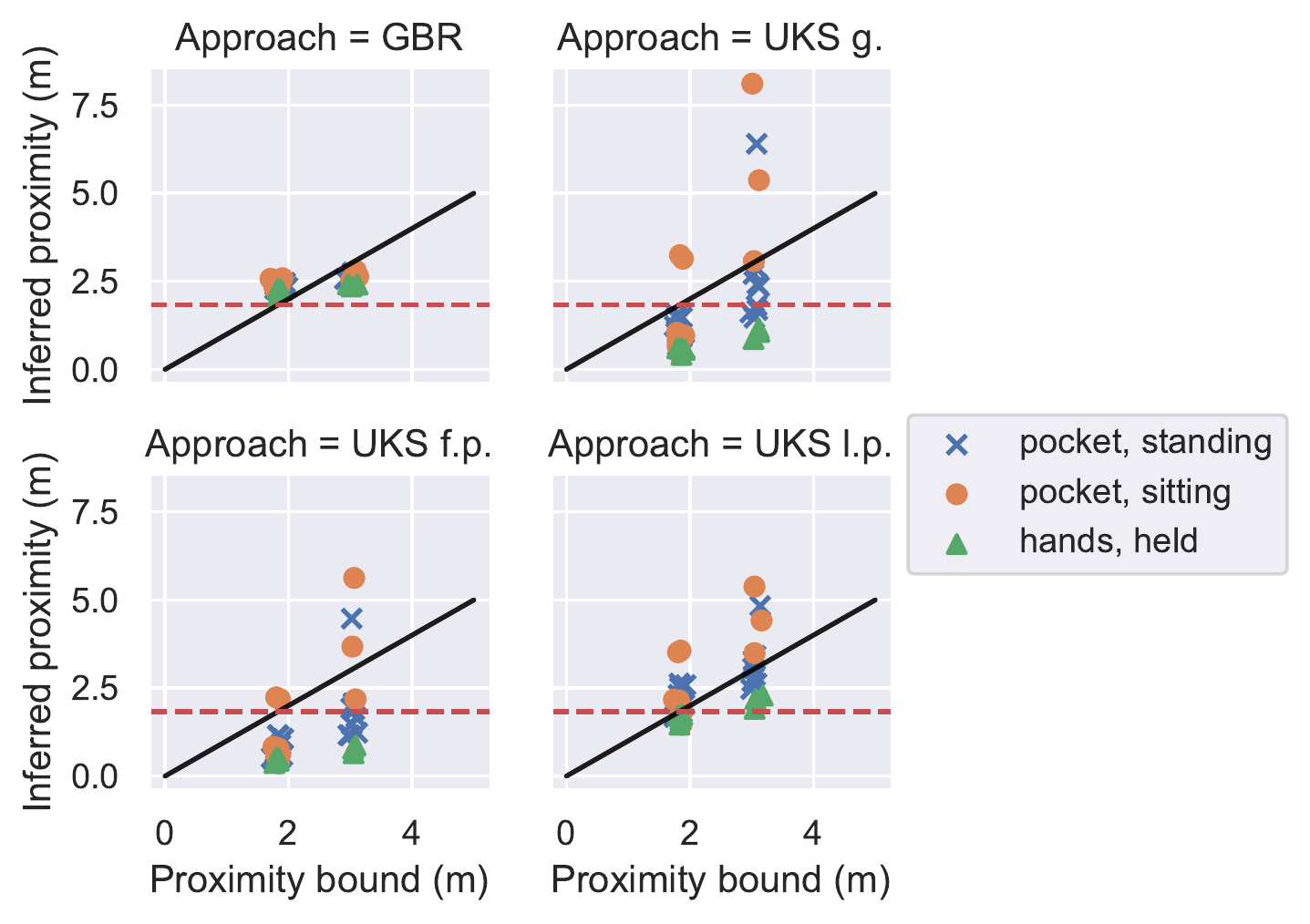} 
  \end{subfigure}
  \caption{Left: relative risk, i.e. inferred risk minus true risk (bound), for MIT H0H1. H1, H1 means the high-risk scenario with high-risk threshold. H0, H0 is the low-risk scenario with low-risk threshold. H0, H1 is the low-risk scenario with high-risk threshold. Right: inferred proximity against true proximity (bound) for MIT H0H1. The two columns of points (with jitter) are the true bounds for H1 and H0 respectively. The red dashed line is the H1 proximity bound.  GBR is the gradient boosted regressor; \gls{uks} g. is the generative model; \gls{uks} f.r./f.p. are the discriminative models in Equation \ref{eq:shift_model} optimised for risk/proximity. \gls{uks} l.r./l.p. are the equivalent for Equation \ref{eq:log_model}. See text for further details on plot interpretation.}
  \label{fig:h0h1_gaussian}
\end{figure}

\begin{figure}[tb]
  \centering
  \begin{subfigure}{.32\textwidth}
    \includegraphics[width=\textwidth]{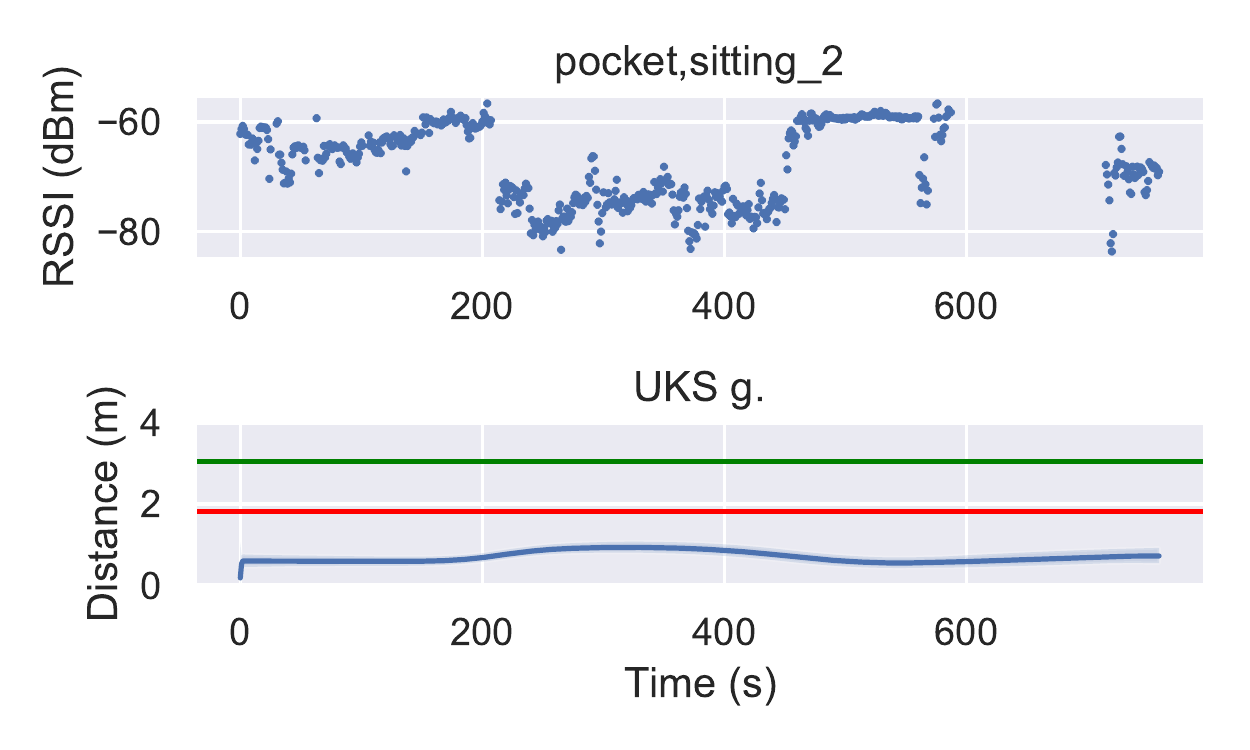} 
  \end{subfigure}
  \begin{subfigure}{.32\textwidth}
    \includegraphics[width=\textwidth]{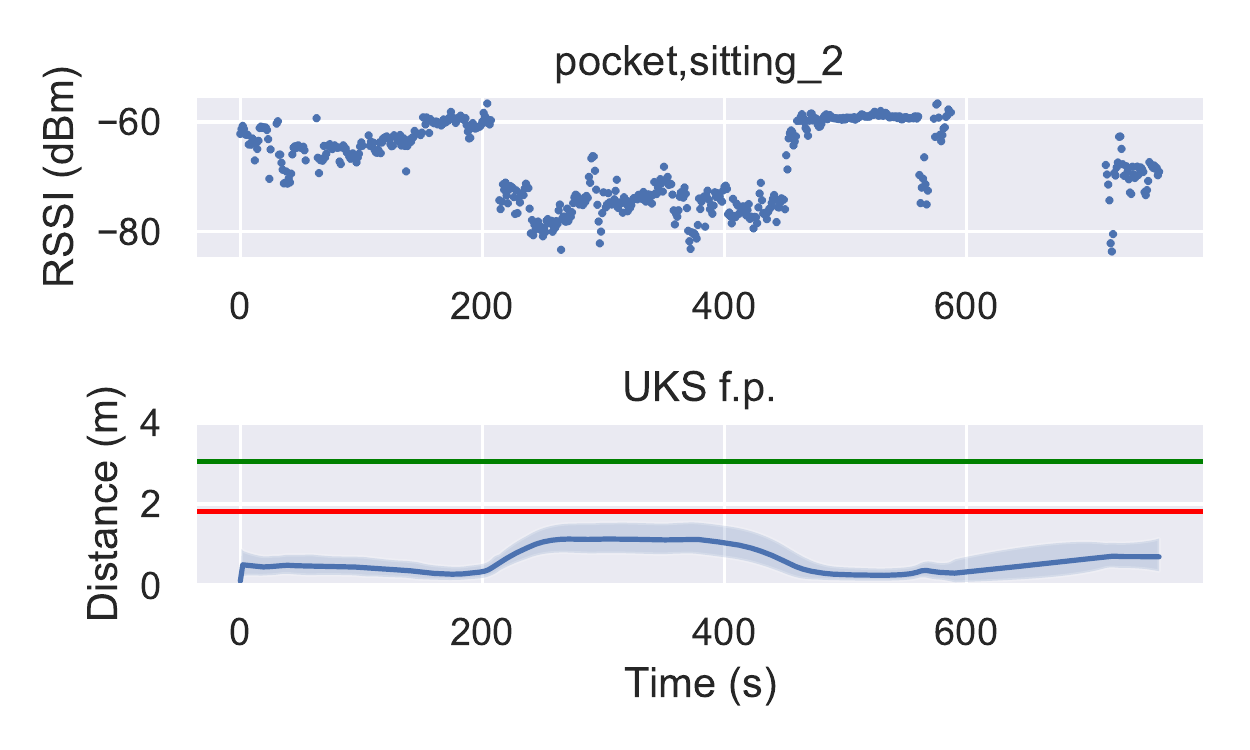} 
  \end{subfigure}
  \begin{subfigure}{.32\textwidth}
    \includegraphics[width=\textwidth]{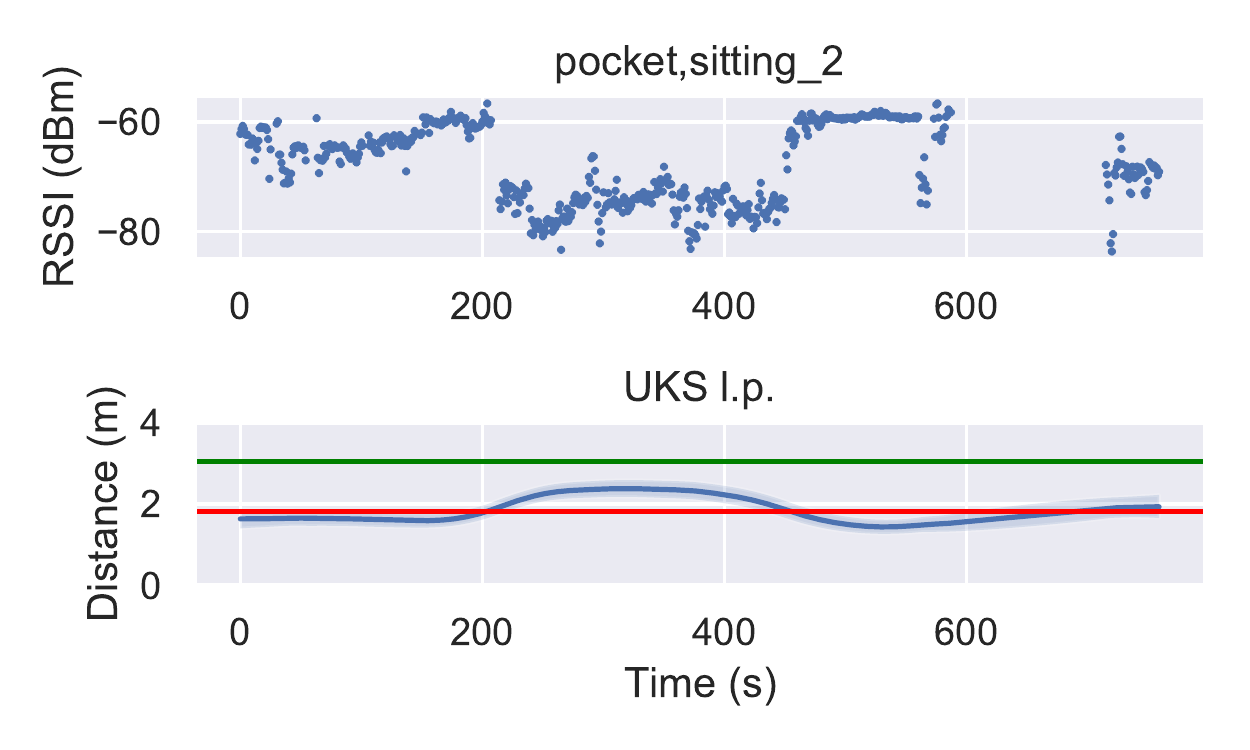} 
  \end{subfigure}
  \begin{subfigure}{.32\textwidth}
    \includegraphics[width=\textwidth]{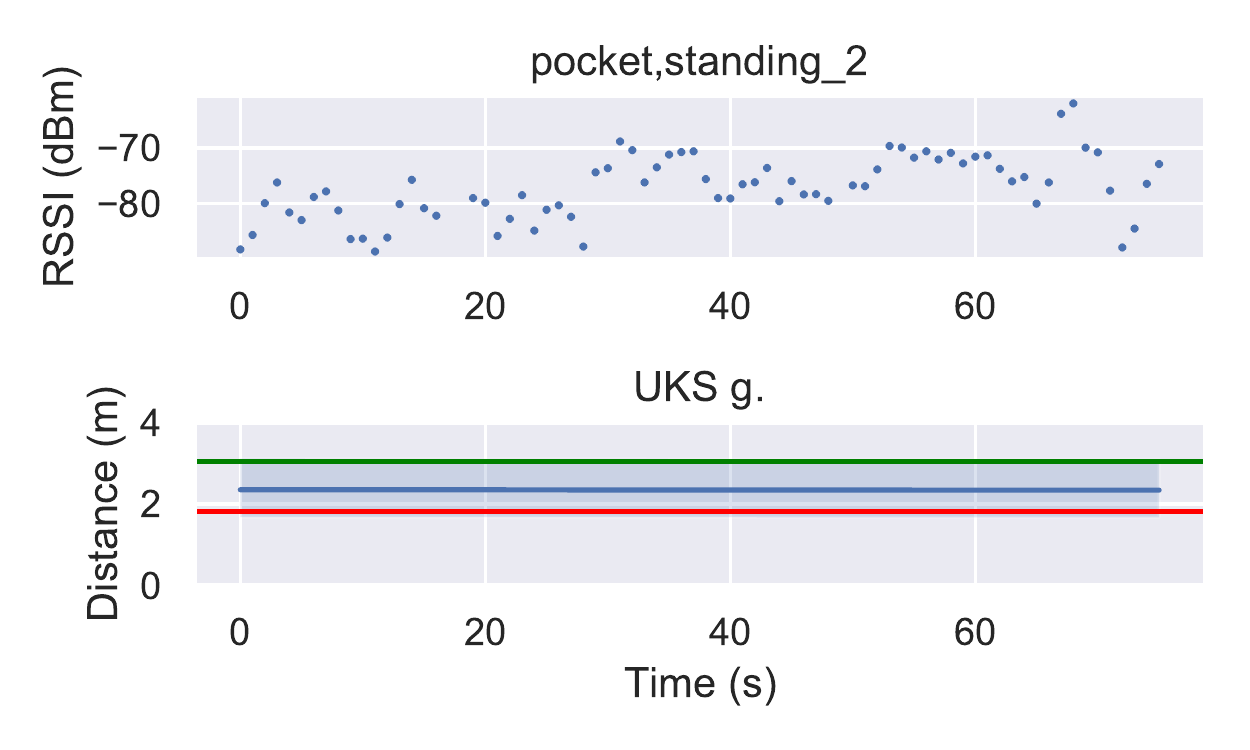} 
  \end{subfigure}
  \begin{subfigure}{.32\textwidth}
    \includegraphics[width=\textwidth]{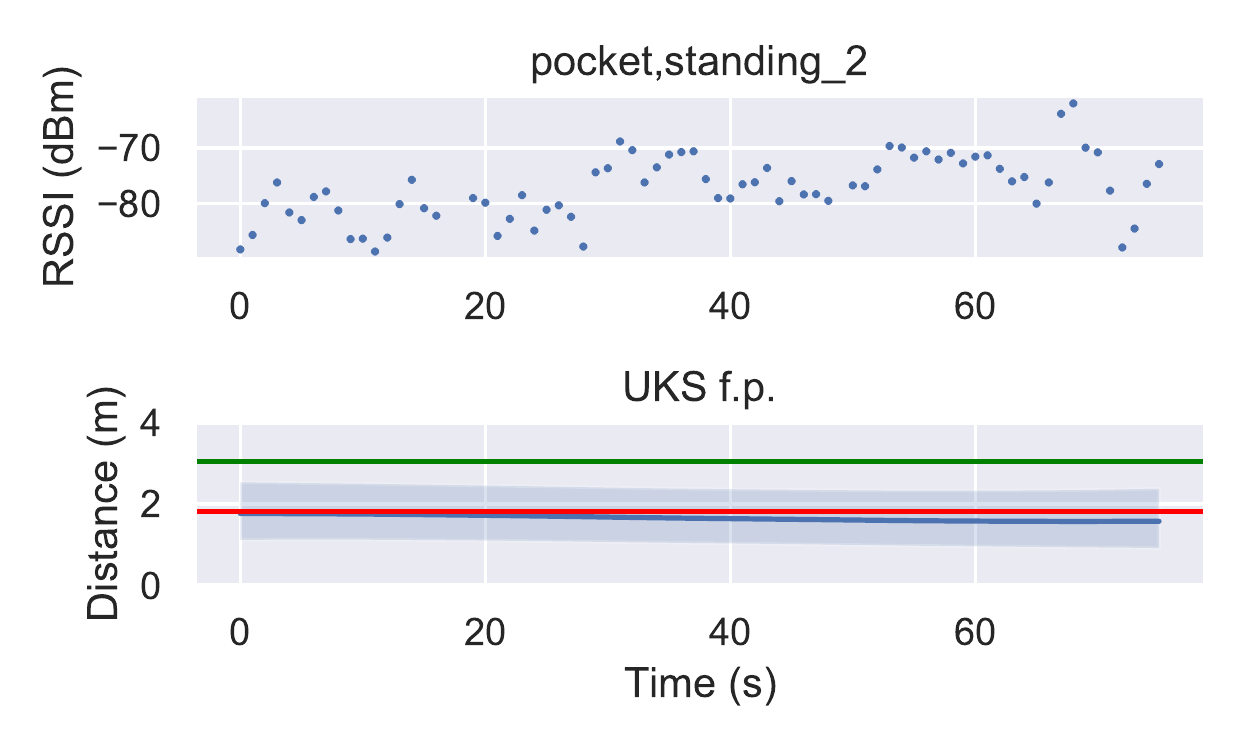} 
  \end{subfigure}
  \begin{subfigure}{.32\textwidth}
    \includegraphics[width=\textwidth]{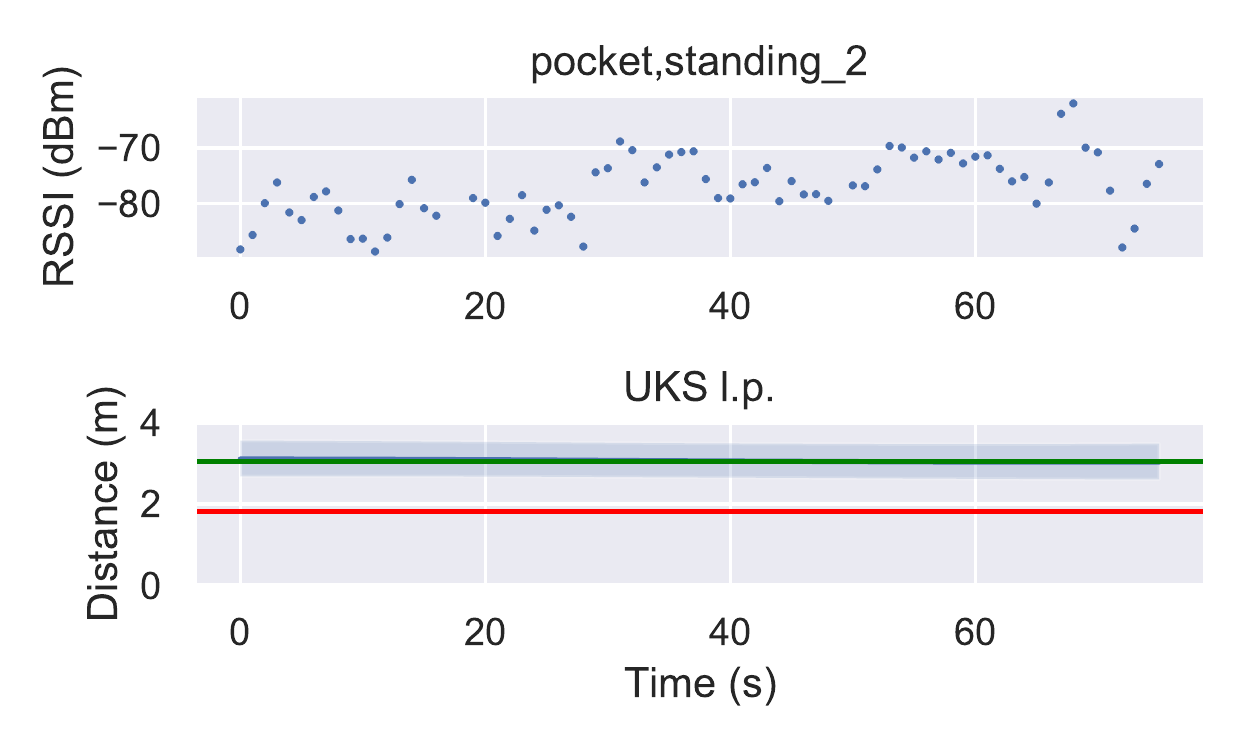} 
  \end{subfigure}
  \caption{Time series of observed \gls{rssi} and \gls{uks} output (mean with $0.05$ and $0.95$ quantiles of a moment-matched gamma distribution) on one H1 example and one H0 example from MIT H0H1. Top row: H1 (high risk scenario); bottom row: H0 (low risk scenario); first column: \gls{uks} with generative model; second column: \gls{uks} with discriminative model (Equation \ref{eq:shift_model}); third column: \gls{uks} with discriminative model (Equation \ref{eq:log_model}). The red horizontal line is the H1 threshold $6$ft, and the green horizontal line is the H0 threshold.}
  \label{fig:h0h1_series_gaussian}
\end{figure}

\begin{figure}[tb]
  \centering
  \includegraphics[width=\textwidth]{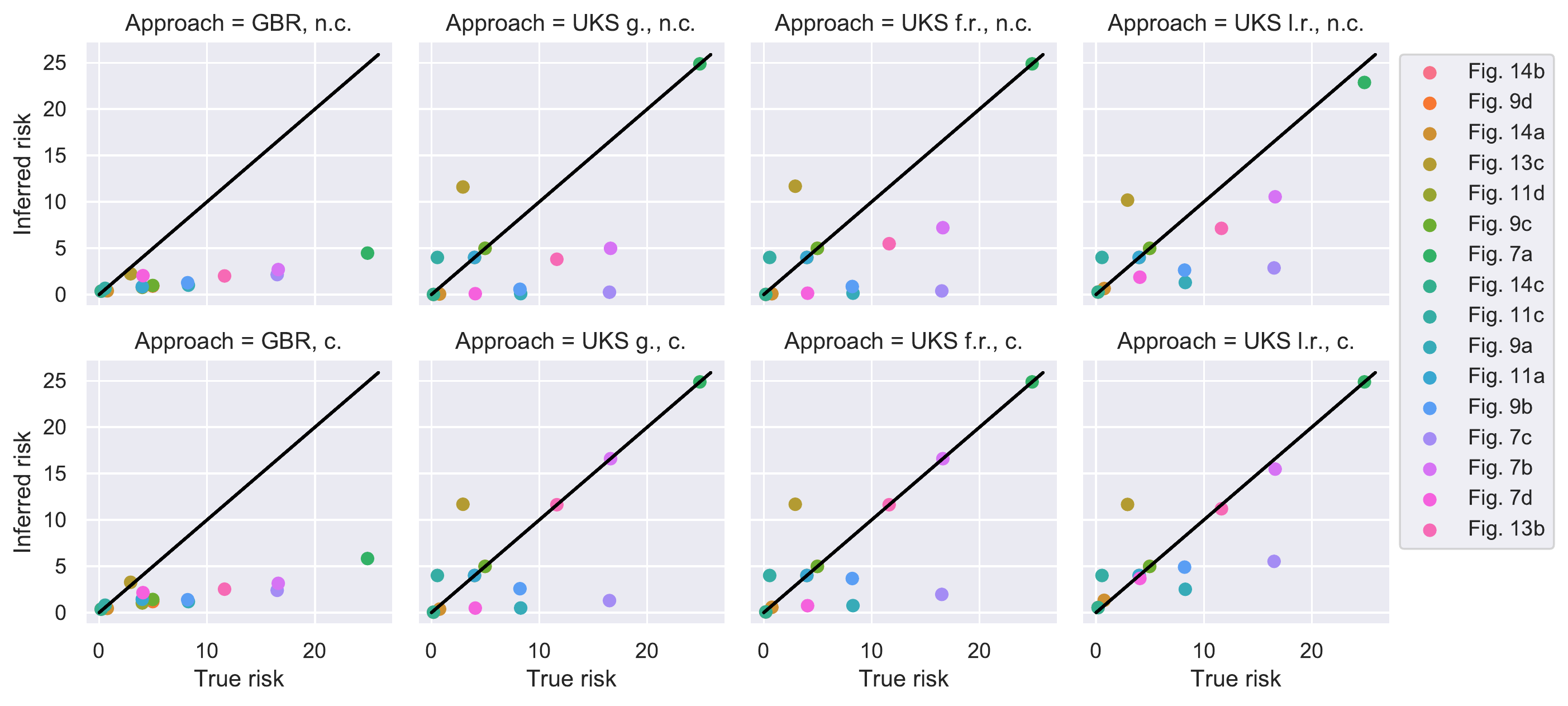}
  \caption{Trinity College data, with inferred risk against true risk. The top row shows the results for raw \gls{rssi} data. The bottom row shows the results when the \gls{rssi} are corrected with the knowledge of mobile device types (Google Pixel 2). GBR is the gradient boosted regressor; \gls{uks} g. is the generative model; \gls{uks} f.r./f.p. are the discriminative models in Equation \ref{eq:shift_model} optimised for risk/proximity. \gls{uks} l.r./l.p. are the equivalent for Equation \ref{eq:log_model}. The plot legend refers to figures in \cite{leith2020coronavirus}. The n.c. and c. refer to ``not corrected'' and ``corrected'' respectively.}
  \label{fig:trinity_gaussian}
\end{figure}

The \gls{roc} \gls{auc} for the approaches are: gradient boosting regressor: 0.5; \gls{uks} g.: 0.728; \gls{uks} f.p.: 0.662; \gls{uks} f.r.: 0.728; \gls{uks} l.p.: 0.769; and \gls{uks} l.r.: 0.6. Figures \ref{fig:random_walk_gaussian}-\ref{fig:trinity_gaussian} show the Figures from the main paper when using the Gaussian model.


\end{document}